# Approaching Landauer's limit in a single superconducting vortex memory bit

Nofar Fridman[1,2,*], Itai Keren[1,2], Avia Noah[3], Yishay Zur[1,2], Martin E. Huber[4], Nadav Katz[1,2,5], Roland Willa[6,7,8], Hadar Steinberg[1,2,**], Yonathan Anahory[1,2,***]

[1] The Racah Institute of Physics, The Hebrew University, Jerusalem, 9190401, Israel
[2] Center for Nanoscience and Nanotechnology, Hebrew University of Jerusalem, Jerusalem, 9190401, Israel
[3] Faculty of Engineering, Ruppin Academic Center, 40250, Emek-Hefer, Monash, Israel
[4] Departments of Physics and Electrical Engineering, University of Colorado Denver, Denver, CO 80217, USA
[5] Qarakal Quantum Ltd., Derech Menachen Begin 144, Tel Aviv 6492102, Israel
[6] Institute of Systems Engineering, School of Engineering, HES-SO Valais-Wallis, rue de l'Industrie 23, 1950 Sion, Switzerland
[7] Institute for Theory of Condensed Matter, Karlsruhe Institute of Technology, 76131 Karlsruhe, Germany
[8] Heidelberger Akademie der Wissenschaften, 69117 Heidelberg, Germany

*Corresponding author Nofar Fridman nofarfri.friedman@mail.huji.ac.il
**Corresponding author Hadar Steinberg hadar.steinberg@mail.huji.ac.il
***Corresponding author Yonathan Anahory yonathan.anahory@mail.huji.ac.il

## Abstract

This work investigates the energy required to manipulate a memory bit using a superconducting vortex confined in a nanoscale double-well pinning potential. In the experiments, various double-well systems subjected to sinusoidal forces is explored, and the effects of both driving force and frequency is analyzed. Results reveal characteristic stochastic resonance behavior for weak drives and adiabatic manipulation at strong, low-frequency drives. Although, our experimental method requires a steady sinusoidal drive incompatible with Landauer's erasure protocol, our numerical analysis demonstrate that the work required for such bit manipulation (erasure) is within an order of magnitude of Landauer's limit ($k_B T \ln 2$) at 4.2 K. These findings open possibilities for creating highly efficient memory elements that could be integrated with cryogenic quantum circuits or function as topological quantum computing qubits; thereby contributing to memory devices with lower energy consumption and mitigating the escalating global energy demand in the computing industry.

## Introduction

Dissipation is crucial in nearly all areas of physical and applied sciences, with its significance further highlighted by Landauer's seminal work, which linked dissipation to information storage in a memory bit.[1] One such memory bit can be encoded in the position of a superconducting vortex[2], which is a candidate for a low-energy consumption classical memory compatible with the cryogenic environment of current quantum computers[3]. Furthermore, a detailed understanding of vortex dissipation impacts quantum technologies more broadly; for example, vortex dynamics contribute to decoherence in superconducting qubits[4,5] and fundamentally affect proposed vortex-based topological quantum information platforms.[6–8]

The problem of many-body vortex matter phases has been extensively studied since the 1990s. Those works probed vortex dynamics involving micrometer displacements at high velocities (km/s) and high vortex concentrations.[9–11] Dissipation was measured by tracking the electric resistance and the out-of-phase components of the magnetic susceptibility.[12] Later, local probes have allowed to zoom in on the single-vortex response under periodic excitation in the low-amplitude limit.[13–19] In some of these studies, local dissipation was detected for AC vortex nucleation[19] and thermal activation near $T_c$.[13,14] Although these works clearly show the presence of dissipation, none of them provides a quantitative estimate.

The low dissipation of single-vortex motion renders its evaluation a formidable experimental challenge. Foltyn *et al.*[20] expel a single vortex from the superconductor over a length of $200 - 300$ nm and show that the resulting dissipation, which gives rise to a local temperature increase of $\sim 0.25$ K, is evaluated to $\sim 4 \times 10^{-19}$ J per operation at $400$ mK. Similar dissipation ($\sim 10^{-18}$ J) was estimated for the creation of a Josephson vortex at $4$ K.[21]

An essential tool for detecting single-vortex motion is the SQUID-on-tip (SOT) microscope, which exhibits high magnetic sensitivity at fine spatial resolution.[17,22–24] SOT probes have also proven to be extremely sensitive local thermometers, achieving a sensitivity of $4 \times 10^{-14}$ W at 4 K.[25] This sensitivity, however, is barely sufficient for detecting the power dissipated by a single vortex driven periodically at our maximum bandwidth (150 kHz), even if the heat released for each period is $4 \times 10^{-19}$ J. It is, hence, unrealistic to detect the dissipation of vortices moving on the nanometer scale using SOT thermometry. Moreover, inferring the heat dissipated by measuring the temperature change requires detailed modeling of the heat flow, including tip-sample and electron-phonon coupling, as well as the Kapitza resistance between different materials.

This work builds on previous studies[13,14,17], to provide a quantitative measure of the dissipation of individual vortices under periodic drive. By measuring the magnetic signal modulation induced by vortex motion, the average work per cycle can be directly assessed without relying on temperature measurements, thereby circumventing the difficulties associated with modeling heat flow. Applying this method, we measure dissipation orders of magnitude below those achieved in previous works, reaching levels below Landauer's limit[1] of $k_B T \ln(2)$.

**Results**

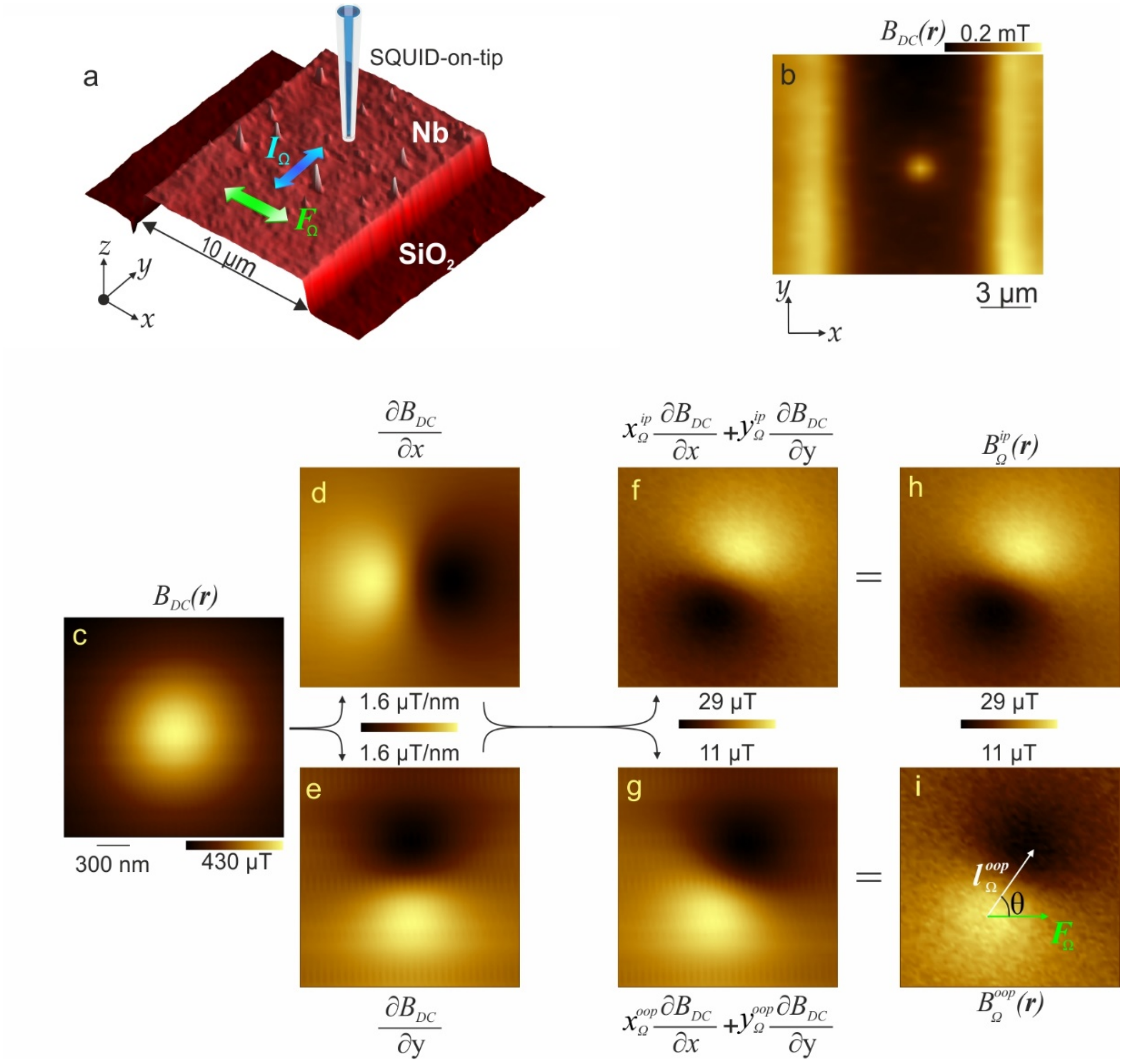

**Figure 1 - Experimental setup.** (**a**) 3D representation of a $16 \times 13$ μm$^2$ AFM image of the 55 nm-thick Nb film patterned into a 10 μm-wide strip. Indicated are the direction of the applied oscillating current $I_\Omega$ (blue double-headed arrow) and the resulting Lorentz force $F_\Omega$ acting on the vortices at the driving frequency $\Omega$ (green double-headed arrow). (**b-e**) SQUID-on-tip images of the $z$ component of the magnetic field. (**b**) DC magnetic image of the out of plane field $B_{DC}(\boldsymbol{r})$ of a single vortex pinned on the Nb strip. (**c**) A zoomed-in image of the measured $B_{DC}(\boldsymbol{r})$ of a vortex. (d-e) Numerical derivatives $\frac{\partial B_{DC}}{\partial x}$ **d** and $\frac{\partial B_{DC}}{\partial y}$ **e**, of the image shown in **c**. (**f**) $x_\Omega^{ip} \frac{\partial B_{DC}}{\partial x} + y_\Omega^{ip} \frac{\partial B_{DC}}{\partial y}$ with $x_\Omega^{ip} = -7.4$ nm and $y_\Omega^{ip} = -15.4$ nm values obtained by fit to **h** (**g**) Same as **f** but for the out-of-phase (oop) component with $x_\Omega^{oop} = 3.2$ nm and $y_\Omega^{oop} = 6.5$ nm values obtained by fit to **i**. (**h-i**) Experimentally measured $B_\Omega^{ip}(\boldsymbol{r})$ **h** and $B_\Omega^{oop}(\boldsymbol{r})$ **i** of the vortex driven by $F_\Omega = 93$ fN at 19 kHz acquired simultaneously with c.

A Nb film of thickness $d = 55$ nm was grown and patterned into a 10 μm-wide bridge (Figure 1a, see Methods). The value of $d$ was chosen to be on the same order as the coherence length $\xi = 12$ nm and the London penetration depth $\lambda = 220$ nm. In this case, the vortex can be regarded as a particle in a random 2D potential landscape (Supplementary Note 1). The sample was then mounted in a scanning SOT microscope to image the out-of-plane component of the magnetic field at 4.2 K (see Methods and Refs. 21, 22). A protocol was established to obtain a single vortex within an area of at least $5 \times 5\ \mu m^2$ as shown in Figure 1b (see Methods). The vortex was subject to a periodic force by applying a current, $I_\Omega = I_0 \cos \Omega t$, along the bridge (blue double-headed arrow in Figure 1a). The resulting Lorentz force is given by $\boldsymbol{F}_\Omega(x,t) = \Phi_0 J_\Omega(x)\hat{x} = F_\Omega \cos \Omega t\, \hat{x}$, where $\Phi_0$ is the magnetic flux quantum and $J_\Omega(x) = \frac{I_\Omega}{\pi\sqrt{(w_b/2)^2-(x)^2}}$ is the sheet current density at a distance $x$ from the center of the bridge of width $w_b$ as calculated in Ref. 26. For example, for a typical current of $I = 0.5$ mA, we obtain $J(0) \approx 31.8$ A/m, and $F_\Omega = 65$ fN $\hat{x}$ (green double-headed arrow in Figure 1a). The relevant force scale was determined experimentally from the observed crossover between linear and nonlinear response as discussed below. The vortex-tip interaction was measured to ~ 7 fN at our scanning distance of 300 nm (Supplementary Note 2), which is comparable to the smallest driving forces $F_\Omega$ and can be partially compensated by a constant current $I_{DC}$.

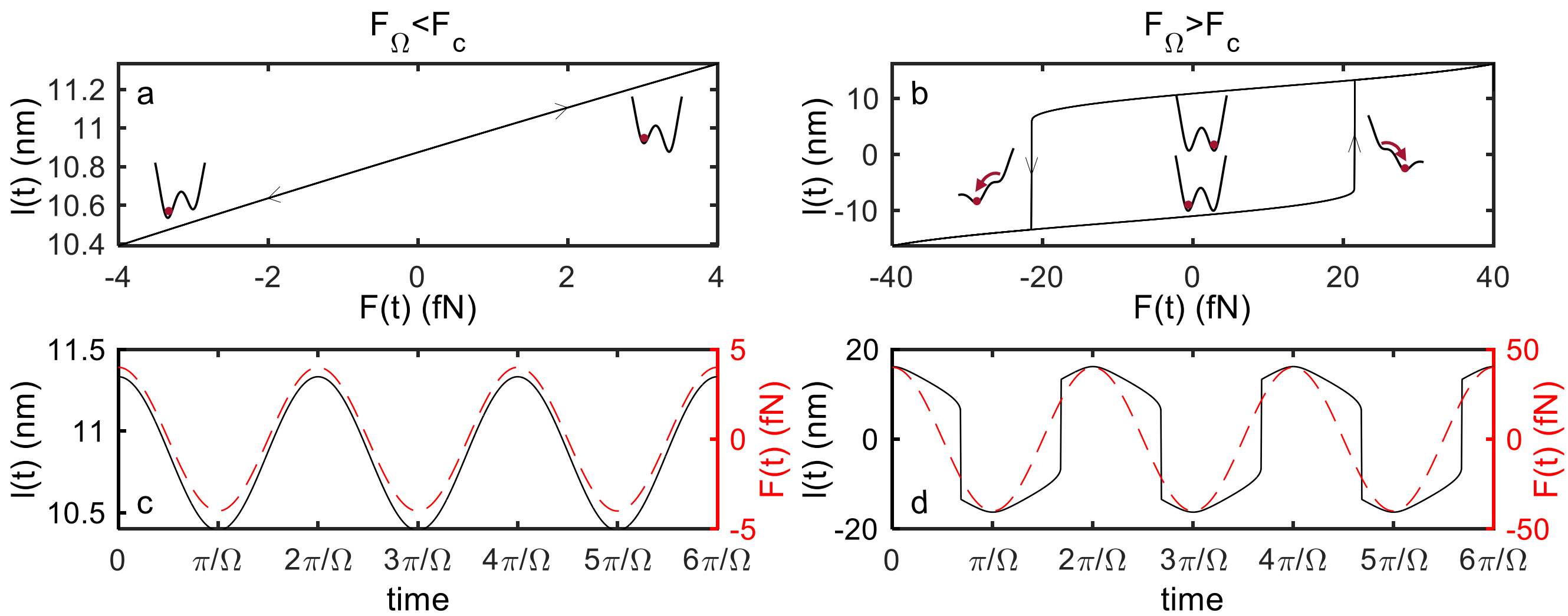

**Figure 2 - One-dimensional toy model of an overdamped double-well potential.** (**a**) Position $l(t)$ vs. applied force $F(t)$ for a vortex oscillating at low amplitude within the left well as depicted in the insets. (**b**) Same as **a**, for larger driving force amplitude $F_\Omega$. The area between the curves corresponds to the vortex dissipation throughout one cycle. (**c**) Time trace of the solution $l(t)$ in **a**. The solution follows the sinusoidal trend of the external force. (**d**) Time trace of the solution given in **b**. Here we observe the apparition of phase delay and higher harmonics. The purpose of this figure is to illustrate the expected physics. None of the curves shown in this figure correspond to measured experimental data or to the Monte Carlo simulations used for quantitative comparison.

The periodic force displaces the vortex away from its equilibrium position, resulting in a time-dependent magnetic field, $B(\boldsymbol{r},t)$. Our methodology relies on measuring the Fourier components of $B(\boldsymbol{r},t)$ with a lock-in amplifier. In general, $B(\boldsymbol{r},t)$ can be expressed as $B_\Omega(\boldsymbol{r},t) = \sum_{n=1}^{\infty} B_{n\Omega}^{ip}(\boldsymbol{r})\cos(n\Omega t) + B_{n\Omega}^{oop}(\boldsymbol{r})\sin(n\Omega t)$, where $B_{n\Omega}^{ip}(\boldsymbol{r})$ and $B_{n\Omega}^{oop}(\boldsymbol{r})$ represent the magnetic field amplitude modulation of the $n$-th harmonic that is in-phase (ip) and out-of-phase (oop) with respect to the phase of the driving current $I_\Omega$, and $\boldsymbol{r}$ is the coordinate of the SOT. The reference phase is set so that the magnetic field caused by the current, according to the Biot-Savart law, is detected solely in the ip signal (see Supplementary Note 3 and Supplementary Figure 1). The scanning SOT microscope simultaneously measures the spatial distributions of the DC magnetic field $B_{DC}(\boldsymbol{r})$ (Figure 1c) and of the 1st harmonic AC field $B_\Omega^{ip}(\boldsymbol{r})$ and $B_\Omega^{oop}(\boldsymbol{r})$ (Figure 1h,i). For small displacements of the vortex, the AC and DC field images are related by $B_\Omega^{ip}(\boldsymbol{r}) = x_\Omega^{ip}\frac{\partial B_{DC}}{\partial x} + y_\Omega^{ip}\frac{\partial B_{DC}}{\partial y}$, where $x_\Omega^{ip}$ and $y_\Omega^{ip}$ are displacement amplitudes at frequency $\Omega$, in-phase with the applied force, along the $x$ and $y$ axis, respectively. Using this relation, we measure the ip and oop displacement amplitude independently as illustrated in Figure 1c-i. The very high sensitivity of the SOT allows us to measure sub-nm displacements.[17] The displacement sensitivity is roughly estimated by dividing the SOT and RMS electronic noise background in a typical $\Delta S_{B_\Omega}$ ~

250 nT by the maximum vortex DC field gradient $\max \frac{\partial B_{DC}}{\partial x} = 0.8\ \mu\mathrm{T/nm}$, which give a displacement sensitivity of $\Delta S_{x_\Omega} \sim 0.3$ nm.

For the rest of this article, we will use polar coordinates $l_\Omega^{ip} = \sqrt{\left(x_\Omega^{ip}\right)^2 + \left(y_\Omega^{ip}\right)^2}$, and $\theta$ is the angle with respect to the Lorentz force as depicted in Figure 1i. The angle between the applied force and the vortex motion arises from the anisotropy of the pinning potential and has been discussed in prior works[17,18].

We calculate the work exerted by the Lorentz force on the vortex $W = \oint \vec{F} \cdot d\vec{l} = \cos\theta \int_0^{2\pi/\Omega} F_\Omega \cos\Omega t \frac{dl}{dt} dt$, where the vortex position is expressed in terms of its Fourier components: $l(t) = \sum_{n=1}^{\infty} l_{n\Omega}^{ip} \cos n\Omega t + l_{n\Omega}^{oop} \sin n\Omega t$. Substituting $\frac{dl}{dt}$ into the integral, all the terms of the Fourier series vanish except the term in $l_\Omega^{oop}$, which results in $W = \pi \cos\theta\, F_\Omega l_\Omega^{oop}$. Given that the average temperature is constant in time, the work exerted on the vortex is equal to the energy dissipated by the vortex.

The dissipation associated with motion in a single-well harmonic potential, as depicted in our toy model Figure 2a,c, is estimated to $W \sim 10^{-25}$ J per cycle (Supplementary Note 4), which is too low to detect experimentally. Thus, we observe $B_\Omega^{ip} \neq 0$ and $B_\Omega^{oop} = 0$, which is expected for the motion of an overdamped particle where the applied force is always equal to the restoring force. However, a double-well potential, as depicted in Figures 2b,d, provides an interesting experimental platform. With our SOT, we resolve – and subsequently study - such a double-well pinning landscape. We search for double-well potentials by exploring distinct equilibrium positions on the sample. By sweeping $I_{DC}$ we tilt the potential landscape moving the average position of the vortex across the sample. For certain values of $I_{DC}$, in addition to the oscillating current $I_\Omega$, a sharp increase of $B_\Omega^{oop}$ is detected. It systematically coincides with the appearance of higher harmonics in $B(\boldsymbol{r}, t)$. This rise in the measured work is interpreted as periodic vortex jumps over an energy barrier that separates two local potential minima (Figure 2b,d). The telegraphic noise measured exclusively when positioning the SOT loop near an undriven vortex confirms this hypothesis (Supplementary Figure 2).

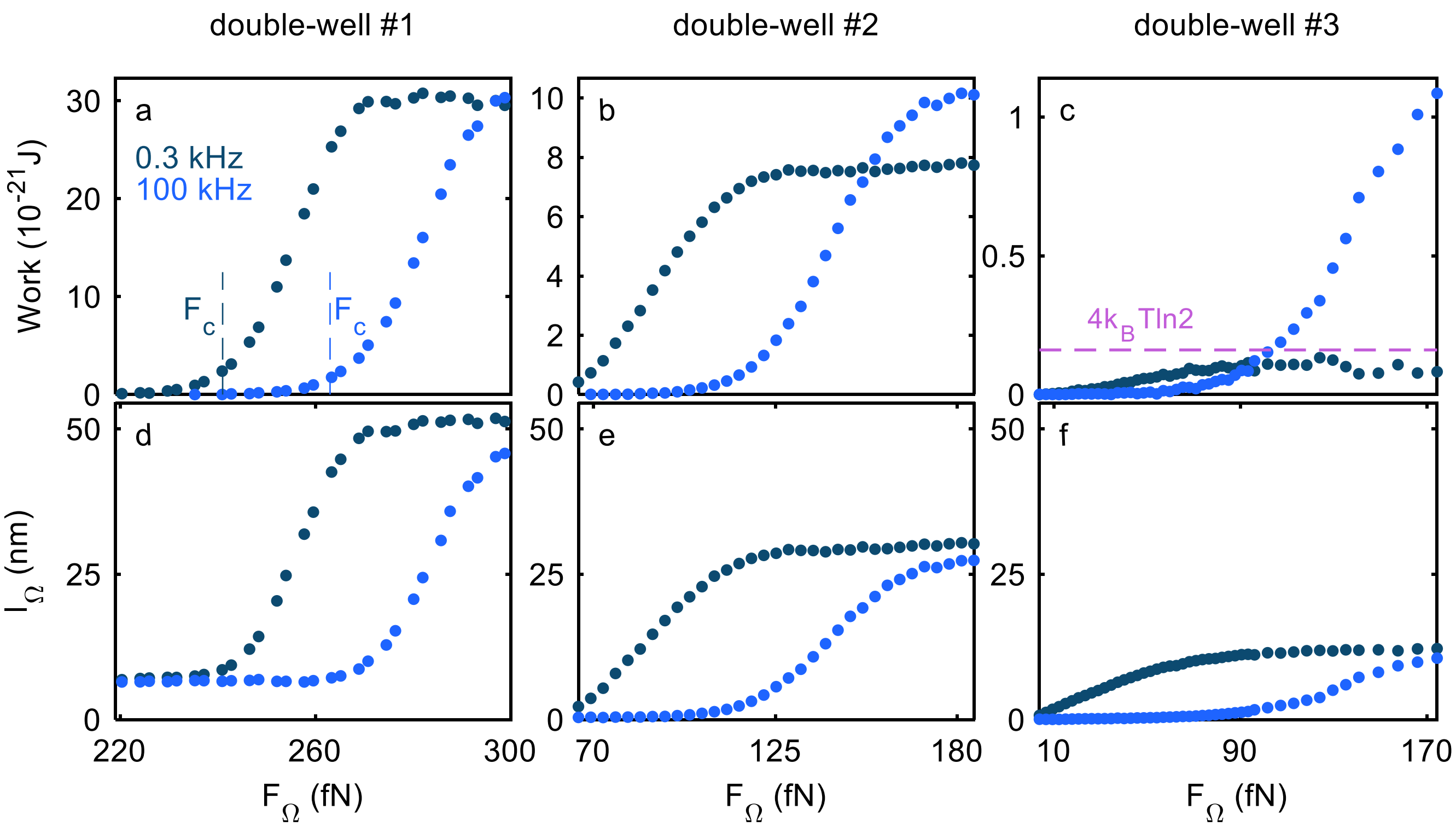

**Figure 3 - Work and displacement as a function of the driving force**. (**a-c**) Average work per period $W$ as a function of the driving force $F_\Omega$ for three double-well potentials located at distinct sample positions. Two driving frequencies $\Omega = 0.3,\ 100$ kHz are plotted in dark and light blue, respectively. (**a**) Vertical dashed curves indicate the critical force $F_c$ at which $W > 0$. (**c**) The pink dashed line marks four times Landauer's limit. (**d-f**) Vortex displacement amplitude $l_\Omega = \sqrt{\left(l_\Omega^{ip}\right)^2 + \left(l_\Omega^{oop}\right)^2}$, as a function of the driving force $F_\Omega$.

Figure 3 shows the work $W$ and displacement $l_\Omega = \sqrt{(l_\Omega^{ip})^2 + (l_\Omega^{oop})^2}$ for three sets of double-well systems as a function of the driving force amplitude $F_\Omega$ in the presence of a fixed $I_{DC}$ for two driving frequencies $\Omega$. We first discuss the common features for $l_\Omega$ seen in all double-well systems presented here. Below a critical driving force $F_\Omega < F_c$, the displacement $l_\Omega$ is homogeneous ($F_\Omega = kl_\Omega$), demonstrating that the vortex oscillation occurs in an overdamped harmonic potential with a spring-constant $k$ (Supplementary Figure 3). As the force reaches $F_\Omega \approx F_c$, $l_\Omega(F_\Omega)$ starts deviating from the harmonic behavior. Notably, for such strong drives, the total real-space displacement amplitude is not equal to $l_\Omega$ due to the presence of higher harmonics. However, our analysis only requires to measure $l_\Omega$ and $W$, which are measured properly even in presence of higher harmonics. Finally, at even higher driving forces $F_\Omega > F_c$, the harmonic behavior is recovered since the barrier can be neglected in the strong drive limit. The value of $I_{DC}$ was chosen to minimize $F_c$ and kept constant for a particular double-well. The general features found here are independent of small $I_{DC}$ changes (see Supplementary Note 5).

We use $W = \pi \cos\theta\, F_\Omega l_\Omega^{oop}$ to extract the average work per cycle (Figure 3a-c). In general, for all investigated double-well sets, for $F_\Omega < F_c$, no work is detected, which is consistent with a vortex trapped in one of the two minima. For stronger drives ($F_\Omega \approx F_c$), the vortex hops occasionally between the two adjacent wells. The vortex hopping to the bottom of the adjacent well, causes a delay in its return and hence a hysteretic force-displacement trajectory (Figure 2b, d). The result is a sharp increase in the expected work. For $F_\Omega \gg F_c$, we find, once more, that the work remains constant, indicating that no significant additional work is performed as the vortex systematically jumps over the barrier in each half-cycle. The work obtained for strong drives varies significantly between different double-well systems. Notably, the work per cycle can be as low as four times Landauer's limit (light purple dashed line in Figure 3c). The origin of the factor four will be discussed below. This has immense potential for low-energy consumption memory devices.

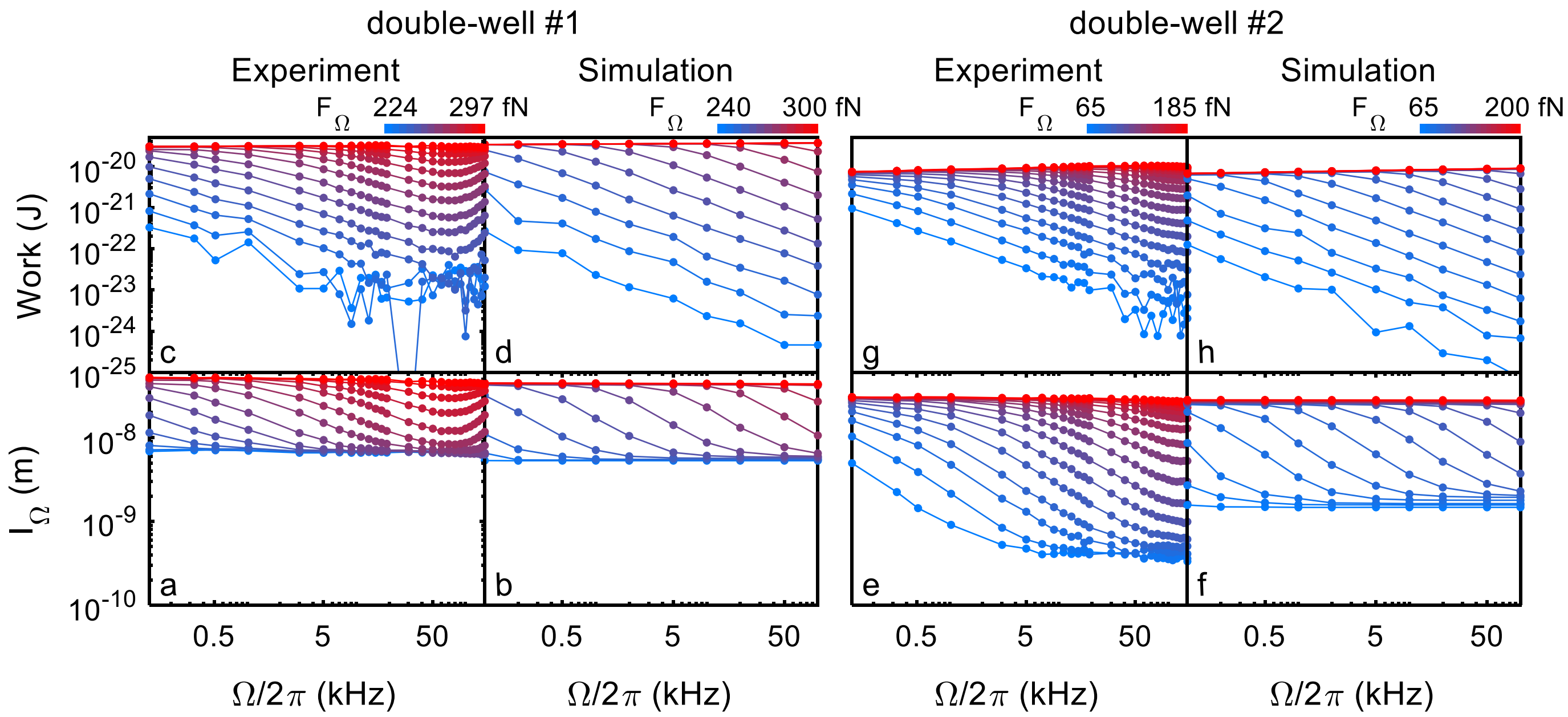

**Figure 4 - Measured and simulated displacement and work as functions of the driving frequency**. (**a-b**), (**e-f**) Displacement $l_\Omega = \sqrt{(l_\Omega^{ip})^2 + (l_\Omega^{oop})^2}$, as a function of the driving frequency $\Omega$ for double-well # 1 **a**,**b,** and double-well #2, **e**,**f**. Experimental data are shown in **a**,**e**; simulations in **b**,**f**. (**c-d**), (**g-h**) Same as **a-b** and **e-f**, but showing the work $W$ as a function of $\Omega$. Experimental data are shown in **c**,**g**; simulations in **d**,**h**. Each curve depicts the measured $l_\Omega$ or $W$ for a given driving force $F_\Omega$. The color of each curve represents $F_\Omega$ as indicated by the respective color bars.

We now turn to the effect of the driving frequency $\Omega$. For the displacement $l_\Omega$, in all three double-well systems reported in Figure 3, an increasing frequency causes an increase in $F_c$. In all these cases, the asymptotic values of $l_\Omega$ at both weak and strong driving forces remain approximately unchanged. The change in $F_c$ can be explained by thermal activation, which plays a greater role at low frequencies. Turning to the dependence of the work $W$ on $\Omega$, we find the same change in $F_c$: For well #1, the work for $F_\Omega > F_c$ remains unchanged within our uncertainty. However, for well #2, we detect a 25% change in the measured work for large $F_\Omega$. That relative difference is much larger for well #3, where work increases by a factor of $\sim 10$ compared to the low driving frequency. We argue here that the evolution of the work with driving force and frequency is a consequence of vortex dynamics in the double-well system at finite temperature. The distinct properties of those double-well systems can be unified within a single model, where the dynamics are explained in terms of the height of the energy barrier $E_b$ between the two wells and the distance between the two minima $2l_0$.

Our model is based on a one-dimensional Monte-Carlo simulation, assuming a double-well potential, as described in detail in the Methods section. A sinusoidal driving force, matching the Lorentz force applied in experiment, tilts the pinning potential. For each time interval $\mathrm{d}t$, the total potential $U_{tot} = U_{pin} + F_\Omega(t)l(t)$ is calculated. The vortex position $l(t)$ always corresponds to a local minimum point of $U_{tot}$ at that time, which means that we ignore the inertia term (overdamped regime). We assume that the typical hopping rate is Kramer's rate $r_K = r_0 e^{-E_b/kT}$, where $r_0$ is the trial frequency. If the vortex jumps over the barrier, we consider instantaneous vortex energy relaxation to align with the potential energy at the new location, implying energy dissipation. We assume that a constant temperature steady state is reached, so the dissipation of energy from the vortex to the bath is equal in magnitude to the work exerted upon it. The average work and displacement per cycle are calculated to be compared with the measured work.

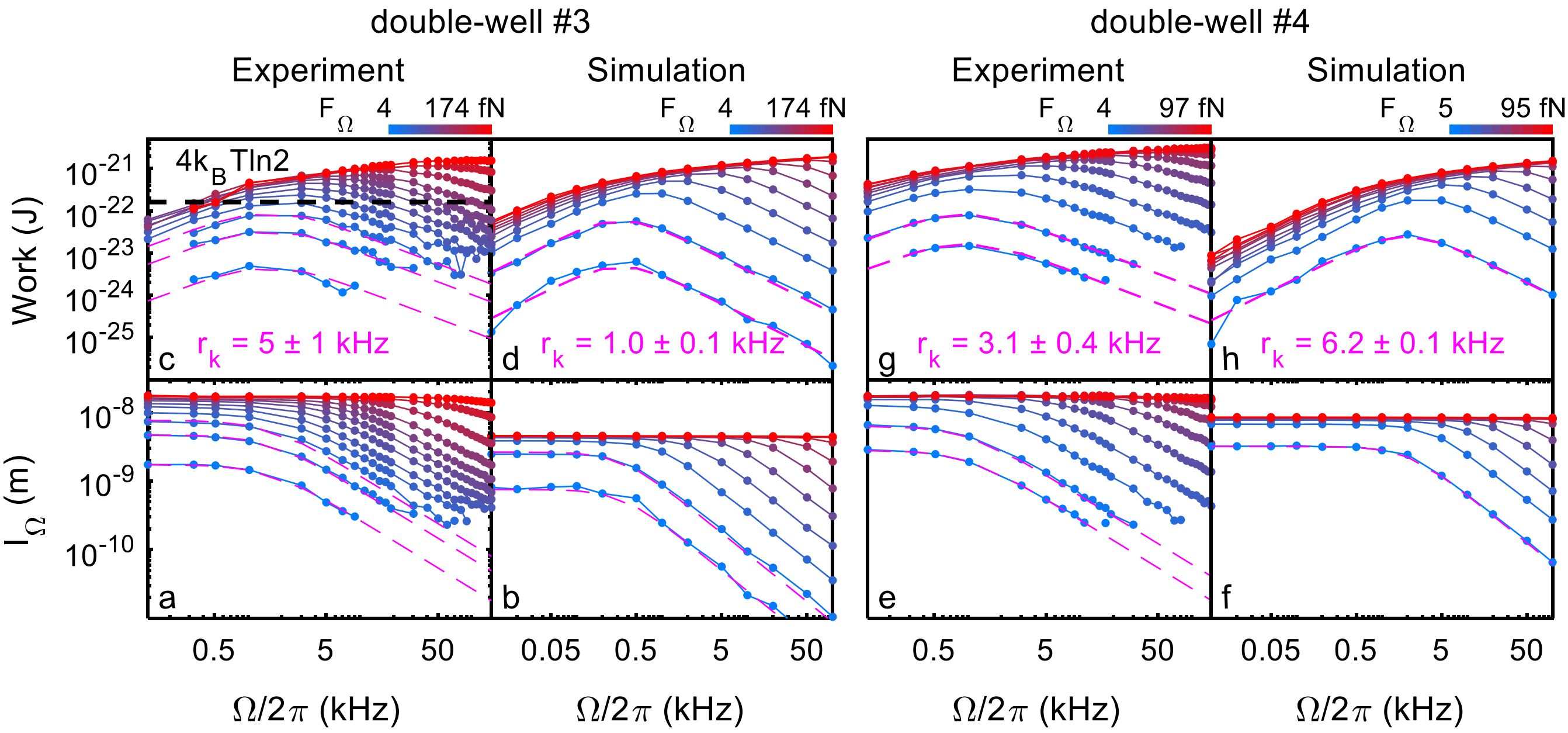

**Figure 5 – Measured and simulated vortex displacement and work for a small energy barrier.** (**a-b**), (**e-f**) Displacement $l_\Omega = \sqrt{\left(l_\Omega^{ip}\right)^2 + \left(l_\Omega^{oop}\right)^2}$, as a function of the driving frequency $\Omega$ for double-well # 3 **a**,**b** and double-well #4 **e**,**f**. Experimental data are shown in **a**,**e**; simulations in **b**,**f**. (**c-d**), (**g-h**) Same as **a-b** and **e-f**, but showing the work $W$ as a function of $\Omega$. Experimental data are shown in **c**,**g**; simulations in **d,h**. Each curve represents the measured $l_\Omega$ or $W$ for a given driving force $F_\Omega$. The color of each curve represents $F_\Omega$ as indicated by the respective color bars. The pink dashed curves correspond to equations (1) and (2) using $r_k$ values specified in the plot legend and $l_0$ values listed in supplementary table 2. The black dashed line marks four times Landauer's limit. Data points resulting from images that did not contain features above the noise level are not shown for clarity.

To test the validity of the model, in Figure 4, we map $W$ and $l_\Omega$ as a function of $\Omega$, given distinct driving force amplitude $F_\Omega$ for the same double-well sets #1 and #2, presented above in Figure 3. For double-well #1, which is the most dissipative and has the largest $F_c$, the vortex displacement $l_\Omega$ is frequency independent in both weak (blue curve) and strong (red curve) drive limits (Fig. 3a). Intuitively, for weak drives, the vortex is trapped in one of the two harmonic potential wells where no frequency dependence is expected (see Supplementary Figure 4 and Note 5). For strong drives, the vortex hops systematically between the two wells, and the energy barrier $E_b$ can be neglected, resulting in a broader harmonic potential with frequency-independent work (Figure 4c). The intermediate regime, where $l_\Omega$ and $W$ scale as $1/\Omega$ over most of the frequency range, is where stochastic effects become significant. Notably, our simulations capture all the features discussed above except for the experimental noise, which becomes visible in the low-dissipation regime ($W \lesssim 10^{-23}$ W). This noise originates from the finite field sensitivity of the SOT and its readout electronics. Double-well #2 exhibits the same features, although it was not probed well into the to the weak drive limit. This regime will be discussed in the following section.

Stochastic resonance theory provides a valuable framework for discussing thermal activation between wells.[27] Taking the limit of perturbative drives, where the external force does not alter significantly $E_b$, the average displacement is described analytically as:

$$l_\Omega = \frac{2F_\Omega l_0 r_k}{k_B T\sqrt{4r_k^2+\Omega^2}} \quad (1)$$

The same framework is used to evaluate the work, which is expected to peak when $\Omega \approx r_k$

$$W = \frac{2\pi F_\Omega^2 l_0^2 r_k \Omega}{k_B T\left(4r_k^2+\Omega^2\right)} \quad (2)$$

Our numerical simulations demonstrate that $r_k$ of the undriven vortex ($F_\Omega = 0$) is far too small to be measured: $r_k = 10^{-45}$ and $10^{-6}$ Hz for double-wells #1 and #2, respectively. In contrast, double-wells #3 and #4 present an opportunity to study a system where thermally activated jumps occur in the perturbative drive limit while placing $r_k$ within our experimental bandwidth. As shown in Figure 5, we measure displacement and work starting with driving forces as low as $F_\Omega = 4$ fN and up to the strong drive limit where $F_\Omega > 100$ fN. At each driving force, we have access to 2 frequency regimes: $\Omega \ll r_k$, and $\Omega \gg r_k$. Figure 5 reports measurements on two sets of double-well potentials and their related simulations (double-wells #3 and 4). We find that stochastic physics is evidently in quantitative agreement with the analytic expressions in equations 1 and 2, using only two fitting parameters for all the curves shown in magenta, for a given double-well. We note that for double-well #3, different values of $l_0 = 2.0$ and $3.9 \pm 0.1$ nm were used to fit $W$ and $l_\Omega$, respectively (see all $l_0$, listed in Supplementary Table 2). This small discrepancy is statistically significant but its origin remains unknown, although the fact that it was not observed in six other examples shown in Supplementary Figure 5 suggests that it might be due to a small systematic error. In the strong drive limit where equations 1 and 2 no longer apply, we can still rely on our numerical simulations, which agree well with the data. We can thus conclude that the stochastic resonance model provides a valid description of the dynamics of double-wells #3 and #4.

The average hopping frequency $r_k$ can be independently probed for an undriven vortex. In this measurement, the SOT is held stationary and positioned at one of the lobes of the AC magnetic field image $B_\Omega(\boldsymbol{r})$, where the signal arising from vortex motion is maximal. In this configuration, the measured signal contains an additional telegraphic noise component originating from the stochastic vortex dynamics, as illustrated in the time domain for another double-well potential in Supplementary Figure 2. We analyze this signal in the frequency domain by measuring the noise spectrum $B(\omega)$ at that location (Supplementary Figure 6). The measured spectrum consists of two contributions: (i) the intrinsic background noise of the SOT, characterized independently when the tip is positioned far from the vortex, and (ii) a Lorentzian component associated with telegraphic noise, expected for a two-state stochastic process, with spectral dependence proportional to $1/\left(4r_k^2 + \omega^2\right)$.[27] This contribution reflects the thermally activated transitions of the vortex between the two minima of the double-well potential. By fitting the measured spectra with this expression, we extract hopping rates of $r_k = 7 \pm 2$ kHz and $r_k = 3 \pm 1$ kHz for double wells #3 and #4, respectively. These values are consistent with those obtained independently from fitting the frequency dependence of the work $W(\Omega)$ and the first harmonic displacement $l_\Omega(\Omega)$ using Eqs. (1) and (2) in the weak-drive limit. This agreement provides further evidence for the stochastic nature of the vortex dynamics and strengthens the experimental basis for both the Monte Carlo simulations and the stochastic resonance analysis.

## Discussion

Landauer's limit was approached or reached in systems where micrometer-scale colloidal particles are manipulated by optical tweezers over a length scale of 1 micron[28], in relatively large SQUID structures[29], and in single-molecule magnets[30]. An interesting outcome of Figure 5 is that the work exerted on a vortex can assume values below Landauer's limit. This limit is of fundamental importance in computer science and sets a bound of $k_B T \ln(2)$ for the heat dissipated to irreversibly erase a memory bit.[1] In the following, we discuss how the parameters used to achieve such a low energy consumption manipulation cannot be used for a memory bit and therefore does not violate Landauer's limit. We demonstrate that using the double-wells presented here, the energy required to manipulate a memory bit indeed exceeds that fundamental limit but remains within the same order of magnitude.

The applicability of Landauer's limit in the different regimes of our experiment is outlined in Figure 6a. For weak drives ($F_\Omega l_0 \ll E_b$), the vortex does not systematically hop over the barrier and therefore, Landauer's limit does not apply (Figure 6a represented in blue). The strong drive limit ($F_\Omega l_0 \gg E_b$) is more nuanced: indeed, for such drives, the vortex is systematically shuttled between the two wells. However, this protocol is not initiated in a random state as assumed in Landauer's analysis. Our experimental method, based on Fourier analysis, imposes a continuous periodic drive and is thus incompatible with Landauer's erasure protocol. We emphasize that a true erasure protocol could in principle be implemented with the double-well potentials presented here, but our present measurement scheme would not allow direct measurement of the resulting dissipation.

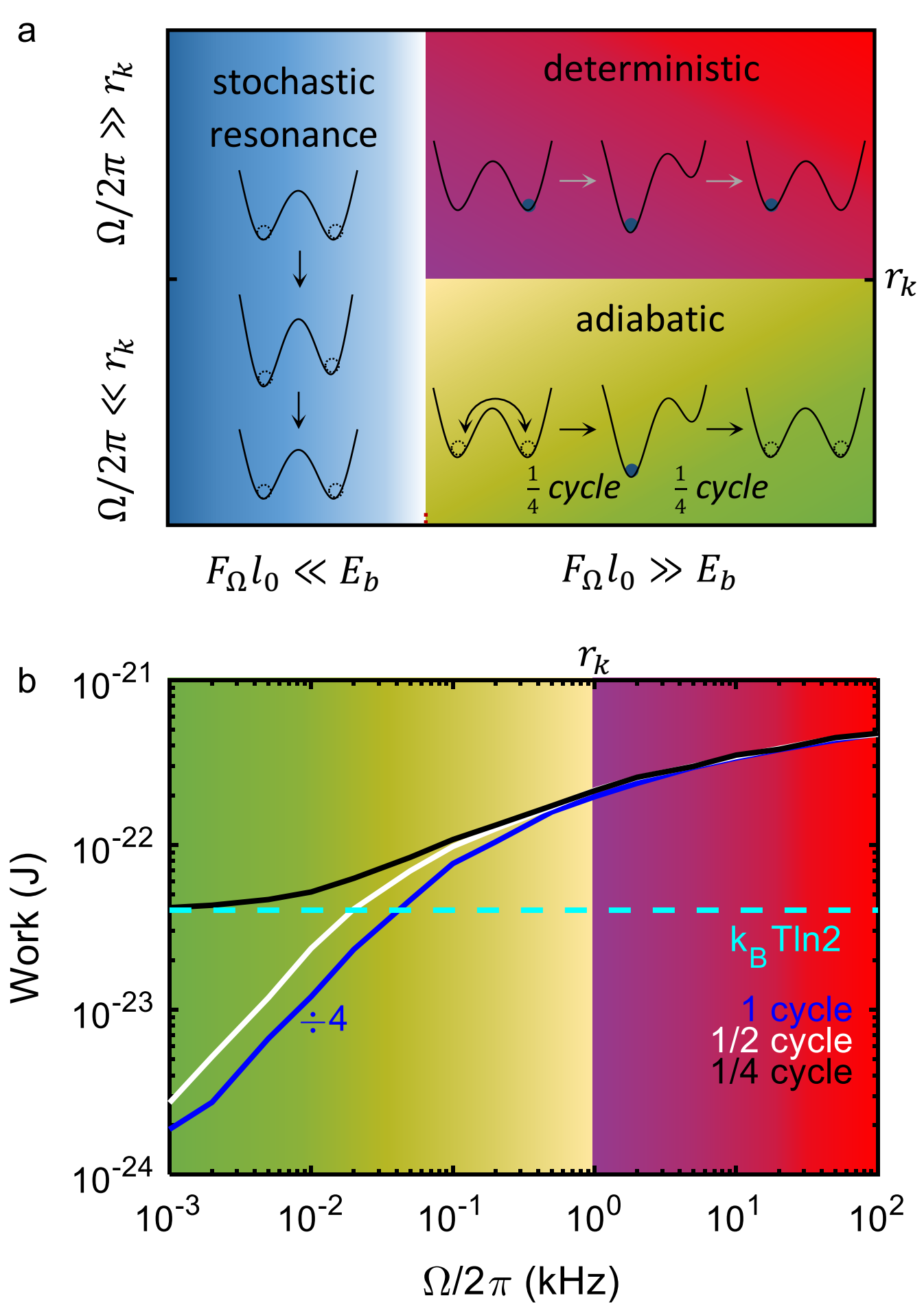

**Figure 6 - Work dissipation in distinct vortex memory protocols: comparison between Landauer's erasure (quarter-cycle), half-cycle, and full-cycle protocols** (**a**) Diagram of the distinct vortex-driving regimes. The region colored in blue is described by the stochastic resonance theory and is valid for perturbative drives $F_\Omega$ at all the driving frequencies $\Omega$. The strong drive limit is divided into two regimes with respect to $\Omega$. Far below Kramer's rate $\Omega \ll r_k = r_0 e^{-E_b/k_B T}$ corresponds to the adiabatic regime where the work per cycle vanishes (colored in green). For $\Omega \gg r_k$, the vortex hops systematically between the wells with no random jumps occurring (colored in red). Insets: schematic representation of the vortex total potential energy $U_{tot}$ at distinct times during the cycle that differ by $1/4$ cycle. Dark blue dots represent the deterministic vortex position, while the dashed line circle depicts the probabilistic vortex position with probability $\approx 1/2$. (**b**) Simulation of the work as a function of $\Omega$ for three different vortex manipulation protocols under strong drive using the parameters that represent double-well # 3. (Black curve) Landauer's erasure protocol, starting from a random state in the untilted double-well potential and ending in the fully tilted potential, which corresponds to a quarter cycle. The plotted curve is the average of 1000 calculations. (White curve) Half-cycle protocol, identical to the Landauer's protocol with an additional quarter cycle, which ends with the potential being untilted as depicted in the inset of panel **a**. (Blue curve) Full-cycle protocol

corresponding to our measurement protocol as shown in Figures 4 and 5. The background color code is the same as in panel **a**, red/green represent the deterministic/adiabatic regime.

This immediately raises the question of how to correctly compare the work measured with our experimental protocol with that required by the Landauer erasure protocol. We provide an answer by running our numerical simulation with a single-shot protocol: the simulation is initialized with an untilted potential, where the vortex is placed randomly in either of the two wells. The work on the vortex in the presence of a periodic force $F_\Omega(t) = F_\Omega \cos \Omega t$ is computed for a quarter period between $\Omega t_{min} = \frac{3\pi}{2}$ (untilted) and $\Omega t_{max} = 2\pi$ (fully tilted). This work calculation is repeated for $1000$ cycles, and the average is taken.

We run this single-shot protocol using the same parameters employed above to simulate the potential landscapes of double-well #3 for the continuous protocol. Figure 6b compares the work resulting from the two protocols. For $\Omega \gg 2\pi r_k$, where stochasticity is not relevant, the work calculated for the single-shot protocol (black curve) is 4 times lower than the work simulated for the continuous protocol (blue curve, shown here with a factor $1/4$). The factor 4 is comprised of a factor 2, related to the complete cycle, where two hopping events occur (back and forth), and another factor 2 related to the initial condition of the vortex. In Landauer's erasure protocol, the initial state of the memory is assumed to be random, with equal probability of being in state 0 or 1. If the system is coincidentally already in the target state, the erasure operation requires no work. As a result, on average, only half of the realization requires hopping over the barrier and thereby contributing to the erasure cost. In contrast, in our experimental protocol, the continuous sinusoidal drive effectively prepares the system in a well-defined initial state prior to each operation. As a result, the vortex must be driven across the barrier in every operation, leading to an average work that is larger by a factor of 2 compared to the Landauer protocol.

For $\Omega \ll 2\pi r_k$, the two situations differ. For the continuous protocol, the work is significantly lower than Landauer's limit, as observed in experiments and simulations. In contrast, the work calculated for the single-shot protocol remains above the limit, confirming the simulation's consistency with Landauer's principle. Notably, in such a regime, the vortex cannot serve as a memory bit since it cannot retain its information on that timescale. Therefore, in our case, a memory bit requires an energy penalty of $\sim 5$ times Landauer's limit at $\Omega/2\pi = 150$ kHz.

To further understand why continuous protocols allow work below Landauer's limit, we run the single-shot model for a half-cycle, allowing the system to return to the untilted potential. Figure 6b white curve highlights that this additional quarter cycle results in work well below Landauer's limit for $\Omega \ll 2\pi r_k$. This low work can be understood by considering that $\Delta S = k_B \ln 2$, which was added to the environment in the first quarter cycle (erasure), is returned to the vortex in the second quarter cycle (randomization), provided that the vortex has sufficient time to randomize ($\Omega \ll 2\pi r_k$). This process is analogous to adiabatic cooling (Figure 6a represented in green).

Finally, our results allow us to evaluate the potential of vortices as low-dissipation memory elements. This could be achieved, for example, by integration with vortex control elements and readout as demonstrated in Ref. 2. Practical memory devices require higher-frequency operation compared to the investigated range of $\Omega$, limited by the bandwidth of our cryostat and SOT microscope. The question is whether higher frequency operation can be achieved without a significant penalty in the required work. In the case of double-wells #1 and #2, where $\Omega$ is at least eight orders of magnitude larger than $r_k$, we see that the work at large drive is frequency independent. It is reasonable to assume that for double-wells #3 and #4, the work per cycle will eventually saturate at a higher frequency. That trend is already visible within our experimental bandwidth. The operational frequency of such a vortex-based memory device would be limited by the shuttling time, which we estimate at 4 ns, corresponding to 250 MHz (Supplementary Note 6). These operational frequencies should be achieved with an energy cost $\approx 10 k_B T \ln 2$.

Based on our results, one can envision using vortices as classical memory bits that interface with quantum computing elements. More exotic applications may involve the use of vortices in topological quantum computing protocols. Consequently, an essential prerequisite is adiabatic manipulation, in which vortices must be shuttled along sophisticated trajectories while retaining their ground state. The type of subgap vortex spectrum could be a determining factor here, which calls for future experiments with topological materials such as $FeTe_xSe_{1-x}$. The experimental methodology we provide here could be used to benchmark vortex manipulation operations and test their adiabaticity.

## Methods

Sample fabrication

The Nb thin film was sputtered in a VST sputtering tool (TFDS-840), with a Kurt J. Lesker 2" Nb target (99.95% purity). The base pressure of the chamber was $3 \times 10^{-7}$ Torr. Sputtering was carried out in a $5 \times 10^{-3}$ Torr Argon for 60 s on a rotating 2" Si wafer at ambient (23 ℃) temperature. The sample was then patterned by direct laser writing photolithography and reactive ion etching with Ar/$SF_6$.

SQUID-on-tip fabrication

The SOT was prepared using a self-aligned three-step process involving thermal deposition of Pb at cryogenic temperatures, as outlined in earlier works[22,23]. The SQUID loop employed in this work has a diameter of approximately 250 nm. This choice of a relatively larger diameter is strategic, as it provides magnetic sensitivity around the zero magnetic field found in the Meissner region of the sample. The larger diameter causes a lower period of the quantum interference pattern of the SQUID critical current versus applied field. This enhances the zero-field magnetic sensitivity, which is essential for measurements at low vortex densities. The tip-to-sample distance was set to 300 nm to be comparable with the tip diameter and to minimize the tip-vortex interactions (see Supplementary Note 2).

Single-vortex trapping protocol

To achieve a low-density vortex distribution, a current pulse exceeding the critical current $I_c$ was applied along the bridge in the $y$-direction for a few seconds, under a near-zero out-of-plane magnetic field $H_z$. The resulting vortex distribution was mapped by imaging the static out-of-plane magnetic field $B_{DC}(\boldsymbol{r})$. This procedure was repeated, with fine magnetic field adjustments $\Delta H_z \lesssim 0.1$ mT, until a single isolated pinned vortex was obtained within an area of at least $5 \times 5$ µm$^2$, as shown in Figure 1b.

Numerical simulations

Our model is based on a one-dimensional numerical simulation where we assume a double-well potential that consists of two identical Lorentzians centered around $\pm x_0$,$U_{pin}(x) = \frac{-U_0}{1+(x+x_0)^2/(\xi_{eff})^2} + \frac{-U_0}{1+(x-x_0)^2/(\xi_{eff})^2}$ ,[17,31] where $\xi_{eff}$ is the well characteristic size and $U_0 \sim 10^{-20}$ J is the depth of each well. To obtain physical pinning sites, we consider $\xi_{eff} \gtrsim \xi = 12$ nm (Supplementary Figure 7). The distance between the two wells, $2l_0 \sim 2x_0$, is set to match approximately the measured $2l_\Omega$ for strong drives. Here we distinguish between $x_0$ the simulation parameter for a given pin and $l_0$ the distance between the two minima resulting from the sum of the two Lorentzian. A sinusoidal driving force, matching the force used at the respective double-well, tilts the pinning potential. For each time interval $\mathrm{d}t$, the total potential $U_{tot} = U_{pin} + F_\Omega(t)l(t)$ is calculated. The vortex position $l(t)$ corresponds to a local minimum point of $U_{tot}$ at that time. We assume that the typical hopping rate is Kramer's rate $r_K = r_0 e^{-E_b/k_BT}$, where $r_0$ is the trial frequency. To achieve this rate, a random number between 0 and 1 is picked. If the number is smaller than $r_K\,dt$, the vortex can overcome the barrier and move to the other local minimum; otherwise, the vortex remains in the same well. If the vortex jumps over the barrier, we assume instant vortex energy relaxation to match the potential energy at the new location. This way, if the vortex jumps from a high/low potential to a low/high potential, the work generated will be positive/negative. The work accumulated over the simulated time is divided by the number of cycles to be compared with the measured work. We assume that the vortex dissipation does not influence the bath temperature. The displacement $l_\Omega$ is calculated by applying a Fourier transform on the position of the vortex $l(t)$ and taking the component at Ω. The trial frequency $r_0$ is assumed to be $10^{12}$ Hz, which roughly agrees with previous values[17,32]. Our general conclusions are largely insensitive to the choice of $r_0$. The simulation parameters are tabulated in the Supplementary Table 1.

## Data availability statement

The data that supports the findings of this study have been deposited in the GitHub database: https://github.com/QIL123/Approaching_Landauers_Limit , DOI: 10.5281/zenodo.21875779

## Code availability statement

The MATLAB scripts that analyze the raw data and reproduce the figures appearing in this paper have been deposited in the GitHub database: https://github.com/QIL123/Approaching_Landauers_Limit. DOI: 10.5281/zenodo.21875779

## Acknowledgements

We want to thank Oded Agam, Oren Raz, Alejandro Silhanek, Joris van den Vondel, Shlomi Kotler, Valla Fatemi, Oded Millo, and Eli Zeldov for fruitful discussions.

## Funding Statement

YA acknowledges the support from the European Research Council (ERC) consolidator grant No. 101124770 (MAJOR). HS acknowledges funding by ISF grant 164/23 and BSF grant 2022122. NK acknowledges support from EU project OpenSuperQPlus and ISF grant 2128/24. YA acknowledges the support from COST Action SUPERQUMAP/CA21144.

### Author Contributions Statement

Y.A., H.S., I.K., and N.F. conceived the experiment.
N.K. and N.F. fabricated the Nb device.
N.F, conducted the scanning SOT measurements.
N.F., I.K. Y.Z., H.S., R.W. and Y.A. analyzed and interpreted the data.
Y.A., N.F. and R.W. conceived the numerical simulations.
Y.A. and A.N. constructed the scanning SOT microscope.
M.E.H. conceived the SOT readout electronics.
Y.A., H.S. and N.F. wrote the article with contributions from all authors.

**Competing Interest Statement**

The authors declare no competing financial interest.

Approaching Landauer's limit in a single superconducting vortex memory bit

Nofar Fridman[1,2,*], Itai Keren[1,2], Avia Noah[3], Yishay Zur[1,2], Martin E. Huber[4], Nadav Katz[1,2,5], Roland Willa[6,7,8], Hadar Steinberg[1,2,**], Yonathan Anahory[1,2,***]

[1] The Racah Institute of Physics, The Hebrew University, Jerusalem, 9190401, Israel
[2] Center for Nanoscience and Nanotechnology, Hebrew University of Jerusalem, Jerusalem, 9190401, Israel
[3] Faculty of Engineering, Ruppin Academic Center, 40250, Emek-Hefer, Monash, Israel
[4] Departments of Physics and Electrical Engineering, University of Colorado Denver, Denver, CO 80217, USA
[5] Qarakal Quantum Ltd., Derech Menachen Begin 144, Tel Aviv 6492102, Israel
[6] Institute of Systems Engineering, School of Engineering, HES-SO Valais-Wallis, rue de l'Industrie 23, 1950 Sion, Switzerland
[7] Institute for Theory of Condensed Matter, Karlsruhe Institute of Technology, 76131 Karlsruhe, Germany
[8] Heidelberger Akademie der Wissenschaften, 69117 Heidelberg, Germany

*Corresponding author Nofar Fridman nofarfri.friedman@mail.huji.ac.il
**Corresponding author Hadar Steinberg hadar.steinberg@mail.huji.ac.il
***Corresponding author Yonathan Anahory yonathan.anahory@mail.huji.ac.il

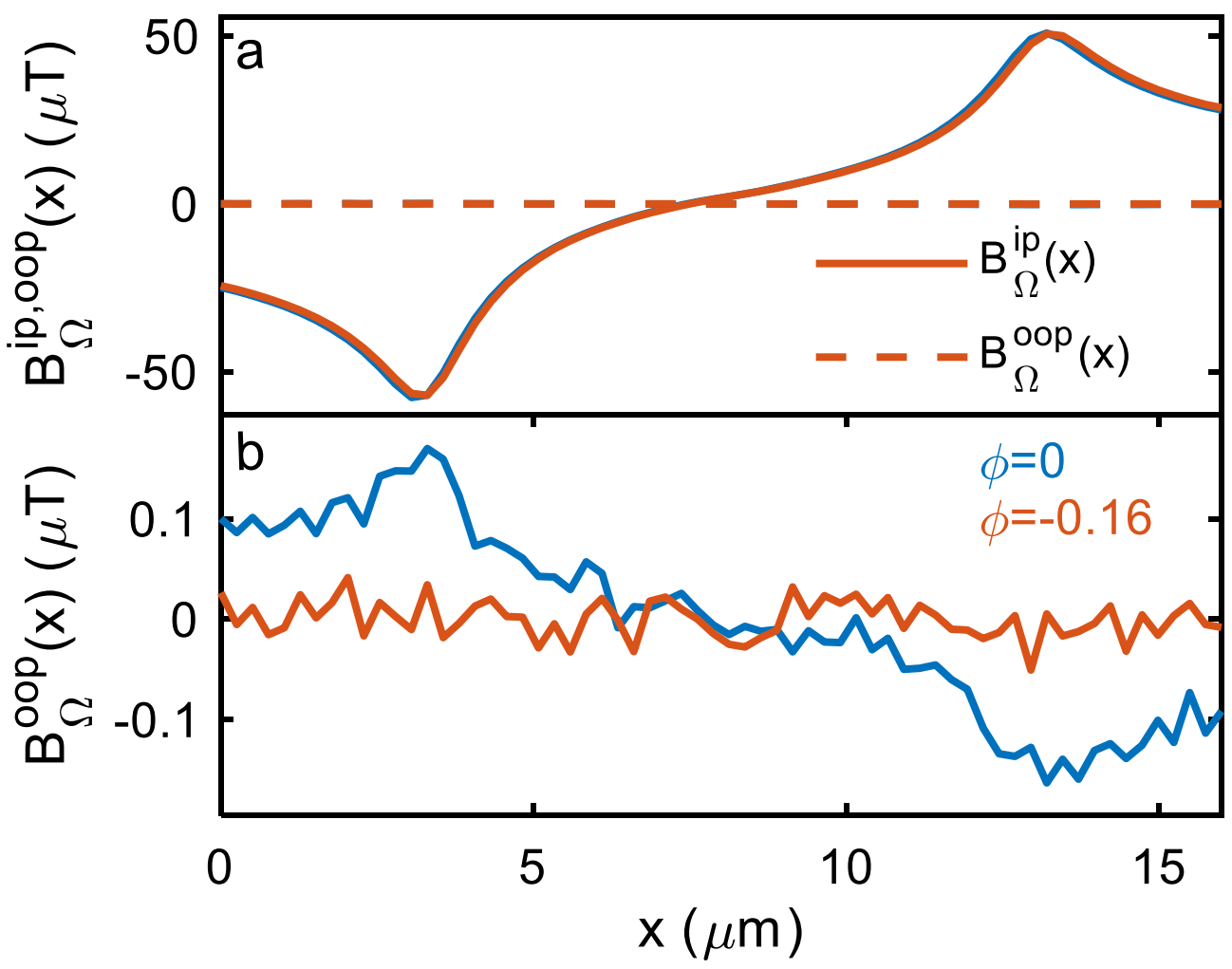

**Supplementary Figure 1 | Phase reference.** (**a**) Solid curves show the measured in-phase (ip) first-harmonic magnetic field $B_{\Omega}^{ip}(x)$ at $\Omega = 7.0$ kHz for $I_{\Omega}(t) = I_{\Omega}\cos(\Omega t)$, $I_{\Omega} = 1.2$ mA acquired with a tip-to sample-distance of 500 nm. The blue and orange curves are nearly overlapping and were obtained with a phase reference of $\phi = 0$ (blue) and $-0.16$ (orange). Dashed curves show the corresponding out-of-phase (oop) first-harmonic magnetic field $B_{\Omega}^{oop}(x)$ representing the signal leakage from $B_{\Omega}^{ip}(x)$ due to imperfect phase reference. (**b**) Same data as in **a**, but zooming in on the dashed curves plotted in **a**. Solid blue and orange curves show $B_{\Omega}^{oop}(x)$ for $\phi = 0$ (before phase correction, blue) and $\phi = -0.16$ (after phase correction, orange).

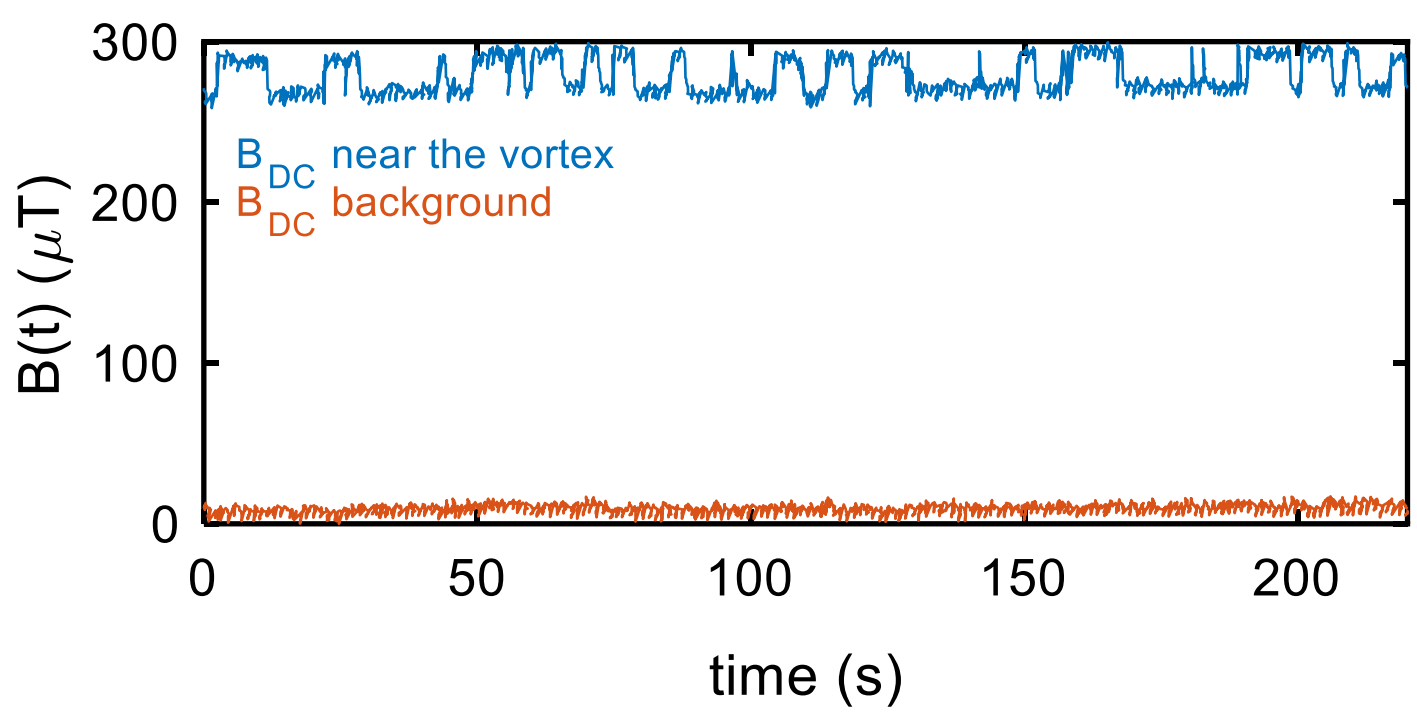

**Supplementary Figure 2 | Telegraphic Noise** Time dependent out-of-plane magnetic field $B(t)$ measured with the SQUID-on-tip (SOT). (SOT) stationed in the vortex vicinity ,(blue). The signal exhibits random jumps between two levels (telegraphic noise), indicating that the vortex is in a double-well potential landscape. (SOT) stationed 1 μm away from the vortex center (orange), no telegraphic noise is detected when the SOT is far from the vortex . Both blue and orange curves were acquired in the absence of an external periodic driving force ($F_{\Omega} = 0$)

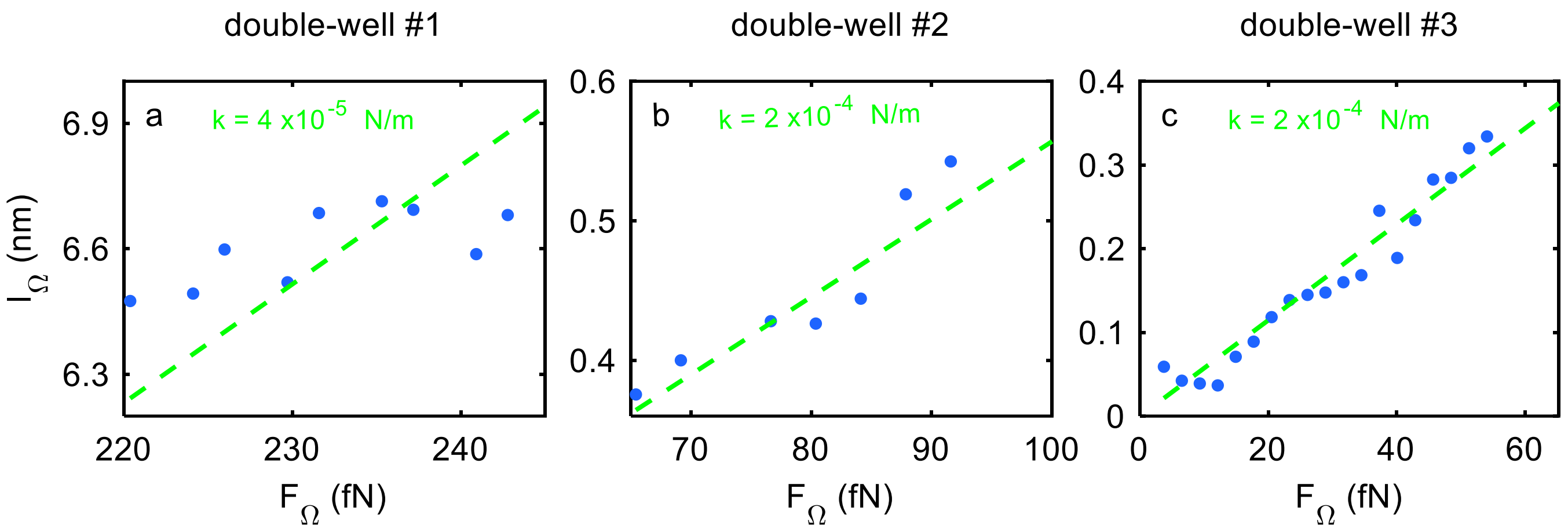

**Supplementary Figure 3 | Displacement $l_\Omega$ vs. driving force amplitude for small driving forces** (**a-c**) Displacement amplitude $l_\Omega = \sqrt{(l_\Omega^{ip})^2 + (l_\Omega^{oop})^2}$, as a function of the driving force amplitude $F_\Omega$ for $F_\Omega < F_c$ where $F_c$ is the critical driving force discussed in Figure 3 of the main text. The data shown here correspond to double-well # 1, 2, and 3 for panel a, b and c, respectively.

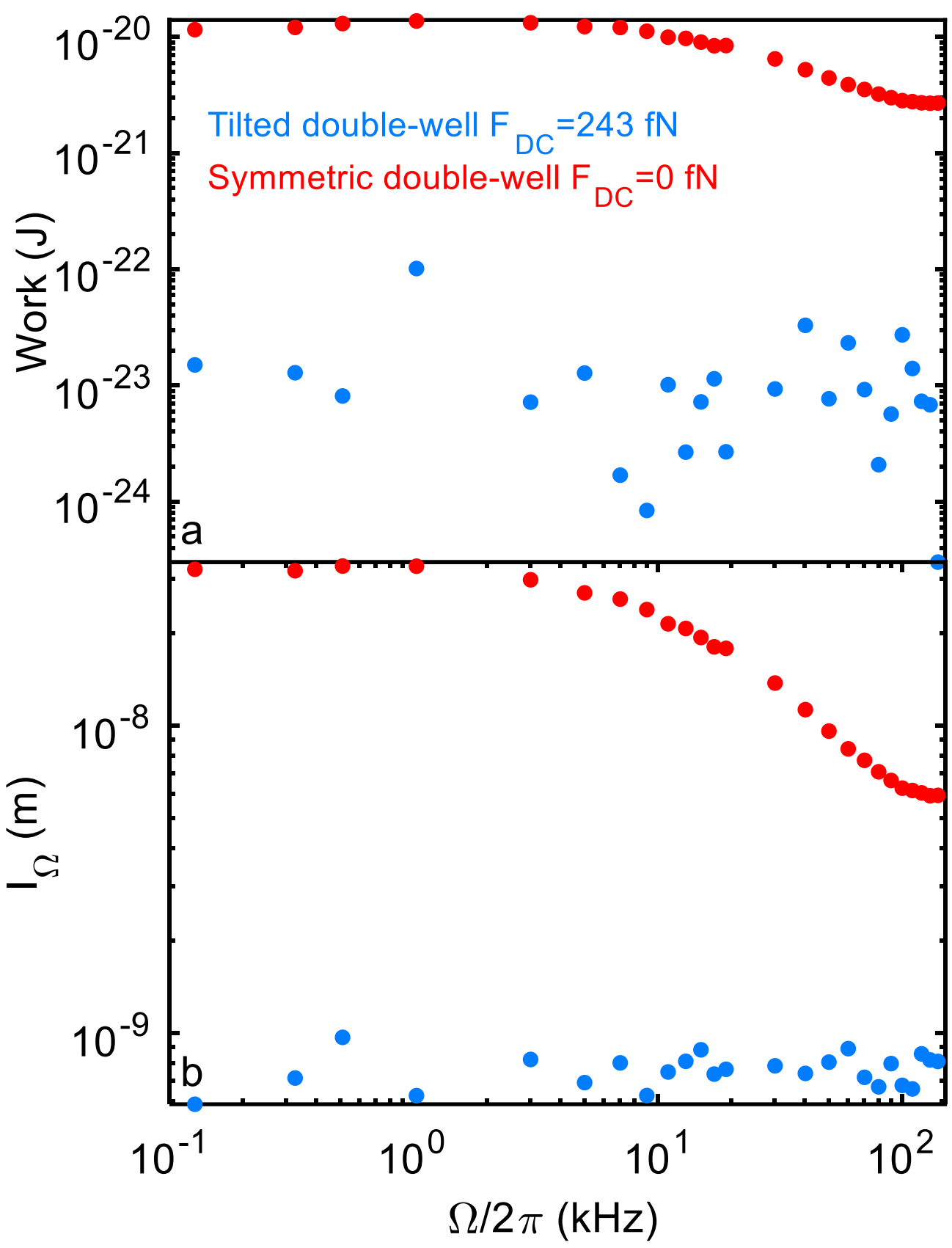

**Supplementary Figure 4 | Work and displacement amplitude for a well in a dissipative balanced state and in a non-dissipative tilted state**. (**a**) Work $W$ as a function of the driving frequency $\Omega$ with a driving force amplitude of $F_\Omega = 188$ fN. The double-well potential is set into two different tilt conditions: no tilt with a static force $F_{DC} \equiv 0$ (red), fully tilted with a static force $F_{DC} = 243$ fN (blue). (**b**) Same as a but showing the displacement amplitude $l_\Omega = \sqrt{(l_\Omega^{ip})^2 + (l_\Omega^{oop})^2}$, as a function of $\Omega$.

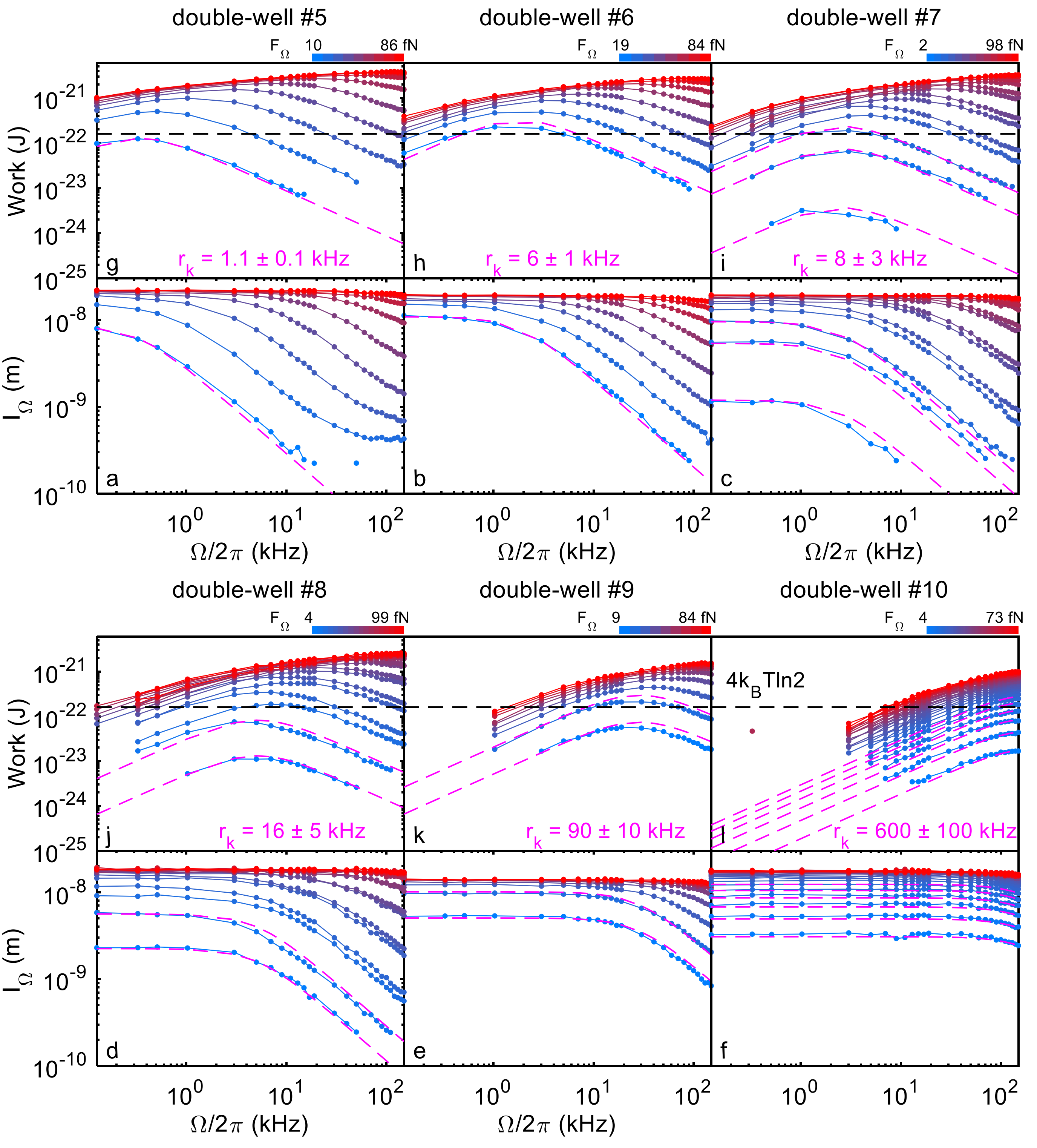

**Supplementary Figure 5 | Measured displacement amplitudes and work as functions of the driving frequency**. (**a**-**f**) Displacement amplitude $l_\Omega = \sqrt{\left(l_\Omega^{ip}\right)^2 + \left(l_\Omega^{oop}\right)^2}$, as a function of the driving frequency $\Omega$ for double wells # 5-10. Magenta dashed curves show the best fits to equation (1) with fitting parameters $r_k$ (Kramer's rate) displayed in the plot and the distance between the wells $2l_0$ (listed in Supplementary Table 2). (**g**-**l**) same as **a**-**f**, but showing the work $W$ as a function of $\Omega$. Pink dashed curves correspond to fits of equation (2) from the main text, using the same $r_k$ and $l_0$ values as **a**-**f**. Black dashed line marks four time Landauer's limit.

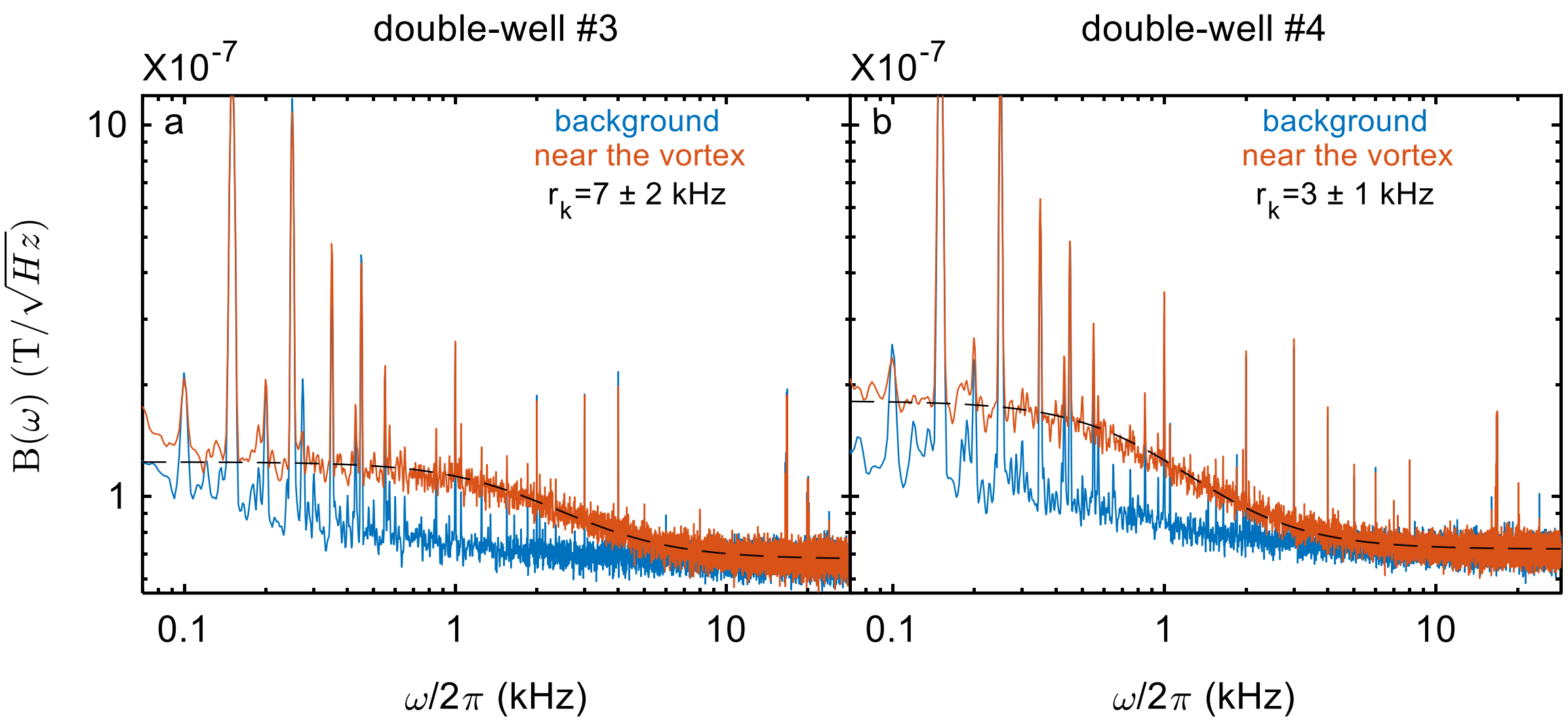

**Supplementary Figure 6 | Power spectra density measured near double-wells #3 and #4**. (**a-b**) (blue curves) Background power spectral density of the magnetic field $B(\omega)$ acquired with the SQUID-on-tip (SOT) at a distance of $\sim 1$ μm from the vortex center and contains solely the experimental setup noise floors. (orange curves) $B(\omega)$ measured with the SOT positioned at one of the lobes seen in $B_\Omega(\boldsymbol{r})$ for double wells #3 **a** and #4 **b** . Measurements were conducted in the absence of an external driving force. (black dashed lines) The experimental data are fitted to a Lorentzian model to extract the characteristic thermally activated hoping rates. $r_k$ = 7 kHz and $r_k$ = 3 kHz are obtained for double-wells #3 and #4, respectively.

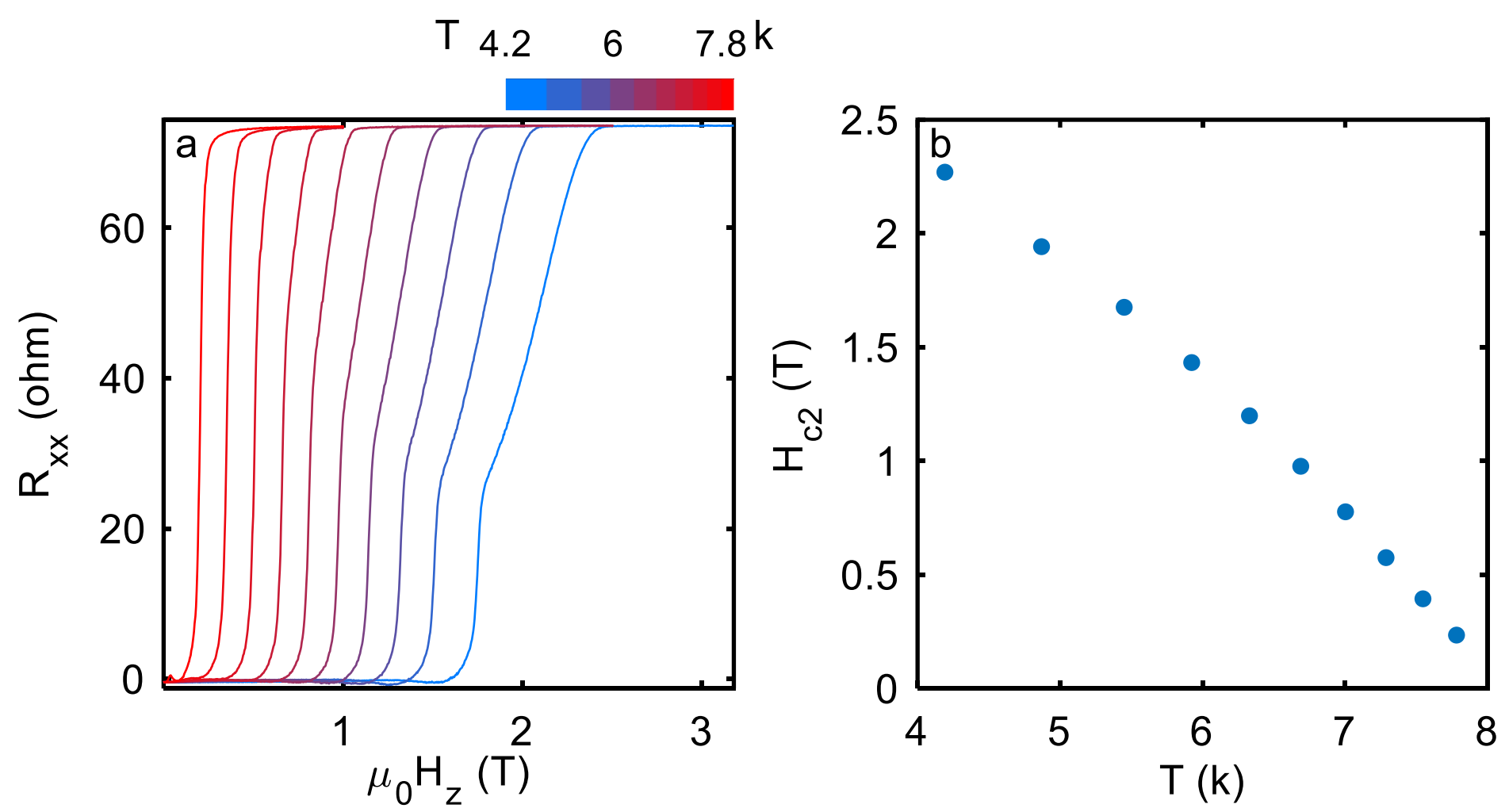

**Supplementary Figure 7 | $H_{c2}$ transport measurements**. (**a**) Longitudinal resistance $R_{xx}$ vs. applied out-of-plane magnetic field $\mu_0 H_z$ for the Nb bridge. Each curve was measured at a distinct temperature as indicated by the color bar. (**b**) Upper critical field $H_{c2}$ as a function of the temperature, determined using the criterion $\frac{r}{r_N} = 0.9$ where $r_N$ is the normal state resistance. The superconducting coherence length at 4.2 K is calculated as $\xi(4.2\ k) = \sqrt{\frac{\phi_0}{2\pi H_{c2}(4.2)}} = 12$ nm.

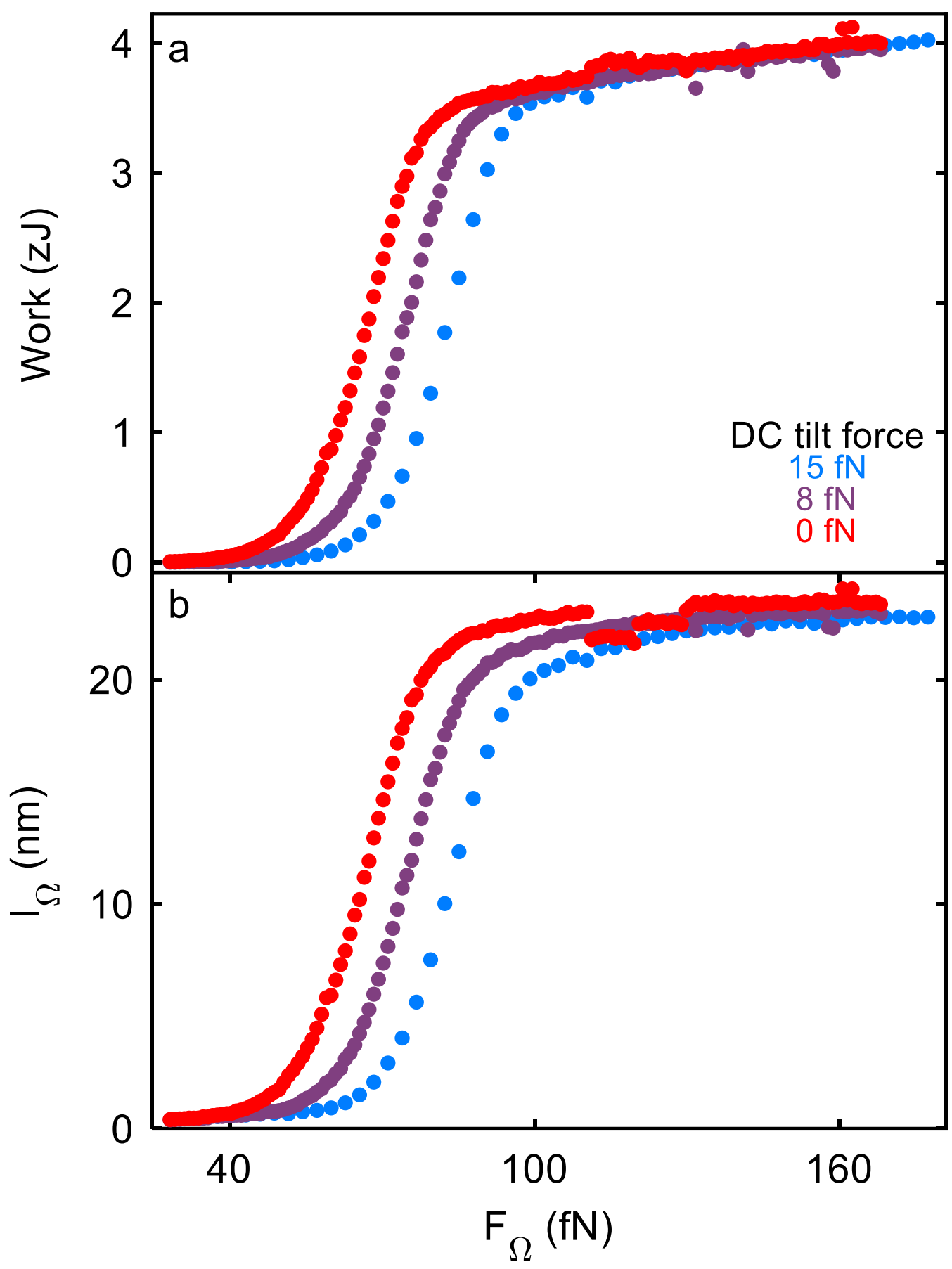

**Supplementary Figure 8 | Work and displacement amplitude as a function of force amplitude** under different potential tilt conditions. (**a**) Work $W$ generated by a sinusoidal force $F_\Omega(t) = F_\Omega \cos \Omega t$, at $\Omega = 7013$ Hz, as a function of the driving force amplitude $F_\Omega$ for a double-well potential under three different tilt conditions: no tilt (red), 8 fN (purple) and 15 fN (blue). (**b**) Displacement amplitude $l_\Omega = \sqrt{(l_\Omega^{ip})^2 + (l_\Omega^{oop})^2}$, as a function of $F_\Omega$, corresponding to the same tilt conditions as in **a**.

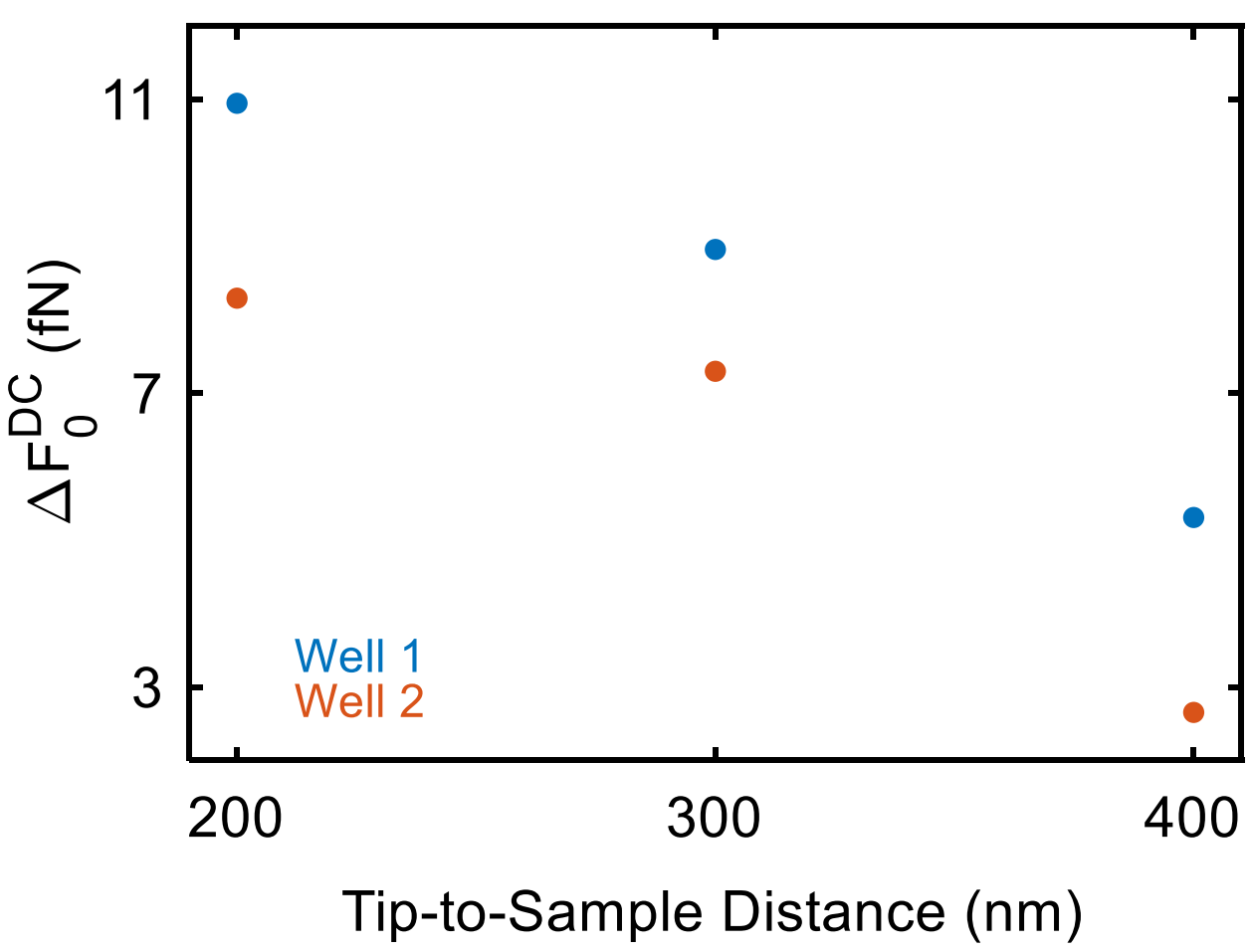

**Supplementary Figure 9 | Interaction force between the Squid-on-tip (SOT) and the vortex**. DC tilt force difference $\Delta F_0^{DC}$ required to balance vortex occupation between the two wells, plotted as a function of tip-to-sample distance. Measurements were performed for both wells: well 1 (blue) and well 2 (orange). For each case, the SOT–vortex interaction was switched from attractive to repulsive by reversing the tip bias, and the tilt force was adjusted until the telegraphic noise indicated equal occupation between the wells. Measurements were performed under a periodic drive $F_\Omega(t) = F_\Omega \cos(\Omega t)$ with $F_\Omega = 60$ fN and $\Omega = 1$ kHz.

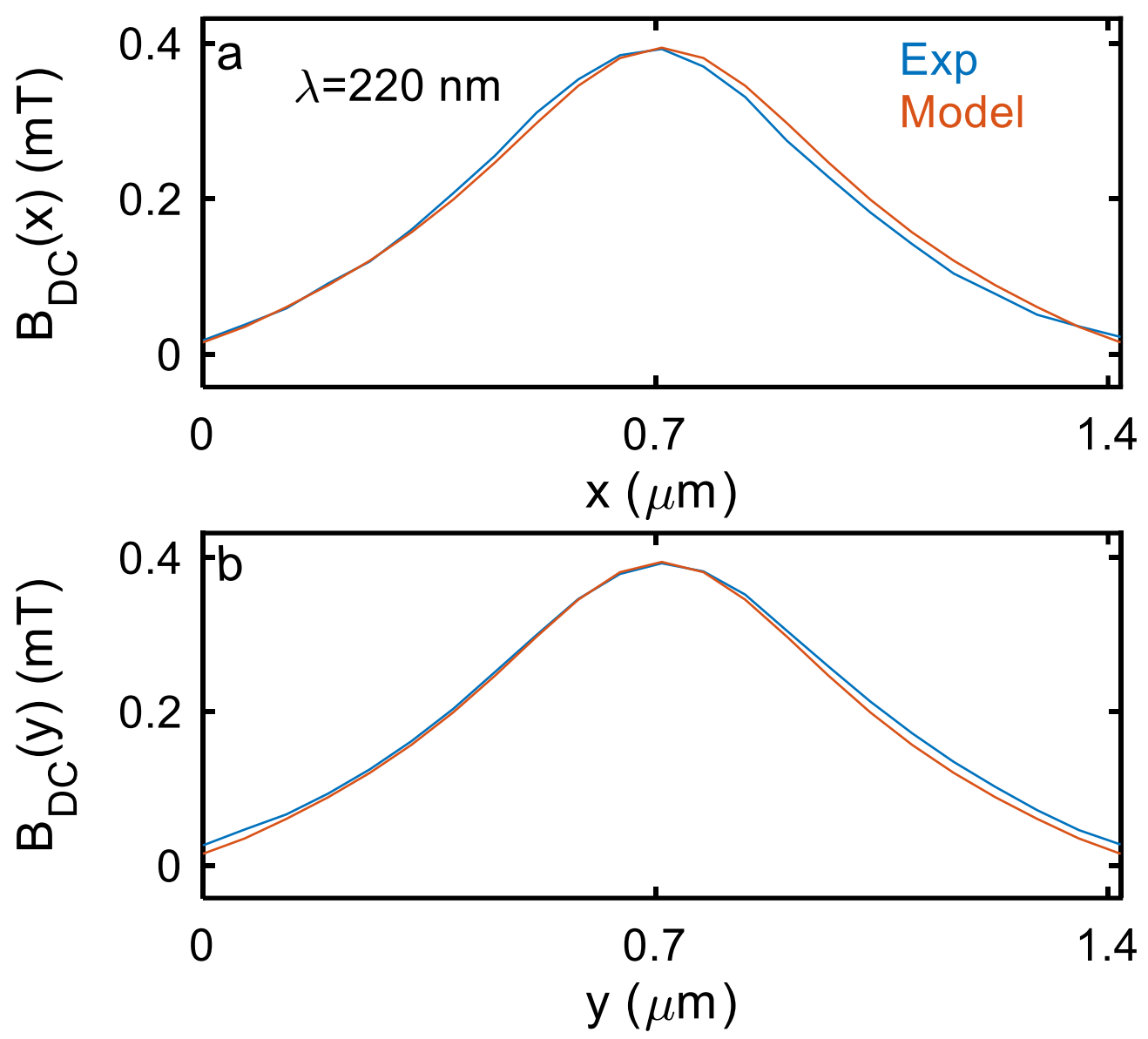

**Supplementary Figure 10 | London penetration length measurement.** (**a-b**) Blue curves represent the out-of-plane magnetic field components $B_{DC}(x)$ and $B_{DC}(y)$ of the vortex profiles along the $x$ axis and $y$ axis, respectively. Orange curves show the best fits using the Pearl model for the corresponding image. The fit yields a Pearl length of $\Lambda = 1.7$ μm. Using the relation $\Lambda = \frac{2\lambda^2}{d}$, with a film thickness of $d = 55\ nm$ we extract a London penetration depth of $\lambda = 220$ nm.

| | double-well #1 | double-well #2 | double-well #3 | double-well #4 |
|---|---|---|---|---|
| $E_b$ ($10^{-21}$J) | $7.5$ | $2.5$ | $1.2$ | $1.1$ |
| $U_0$ ($10^{-21}$J) | $36$ | $16$ | $230$ | $24$ |
| $\xi_e$ (nm) | $36$ | $24$ | $12$ | $12$ |
| $x_0$ (nm) | $36$ | $22$ | $7.5$ | $8.8$ |
| $r_0$ (Hz) | $10^{12}$ | $10^{12}$ | $10^{12}$ | $10^{12}$ |
| $F_\Omega$ (fN) | $240{:}\,300$ | $65{:}\,200$ | $4{:}\,174$ | $5{:}\,95$ |

**Supplementary Table 1 | List of input parameter values employed to simulate each double-well**. $E_b$ is the energy barrier between the two wells for the untilted potential $F_\Omega(t) = 0$. $U_0$, $\xi_{eff}$ and $x_0$ defines the potential shape $U_{pin}(x) = \frac{-U_0}{\left(1+\left(\frac{x+x_0}{\xi_{eff}}\right)^2\right)} - \frac{U_0}{\left(1+\left(\frac{x-x_0}{\xi_{eff}}\right)^2\right)}$. $U_0$ is the potential well depth of the two Lorentzian potentials. $\xi_{eff}$ defines the length scale over which the potential goes back to zero, for a physical potential landscape, it cannot be smaller than $\xi = 12$ nm. $x_0$ defines the distance between the two Lorentzian potentials. $r_0$ is the trial frequency. $F_\Omega$ is the driving force amplitude.

| | Experiment $l_\Omega$ fit | Experiment $W$ fit | Simulation fit | Simulation input |
|---|---|---|---|---|
| well #3 $l_0$ (nm) | $3.9 \pm 0.1$ | $2.0 \pm 0.1$ | $3.3 \pm 0.1$ | $3.45$ |
| well #4 $l_0$ (nm) | $6.6 \pm 0.1$ | | $6.1 \pm 0.1$ | $6.336$ |
| well #5 $l_0$ (nm) | $7.2 \pm 0.2$ | | - | - |
| well #6 $l_0$ (nm) | $5.8 \pm 0.2$ | | - | - |
| well #7 $l_0$ (nm) | $5.9 \pm 0.3$ | | - | - |
| well #8 $l_0$ (nm) | $5.6 \pm 0.1$ | | - | - |
| well #9 $l_0$ (nm) | $6.9 \pm 0.2$ | | - | - |
| well #10 $l_0$ (nm) | $5.9 \pm 0.4$ | | - | - |

**Supplementary Table 2 | $l_0$ obtained by fitting equations #1 and #2 in the stochastic resonance regime.** List of the distances between wells $2l_0$, obtained from fits to Equations (1) and (2) of the main text, for both the experimental and simulated double-wells shown in Figure 5 and in Supplementary Figure 5. For all double-wells except #3, the same $r_k$ and $l_0$ values were used to fit multiple curves corresponding to both the displacement $l_\Omega$, Eq. (1), and the work $W$, Eq. (2). The fits are shown in Figure 5 and Supplementary Figure 5 (in magenta). For double-well #3, $r_k$ value is the same while $l_0$ differed between $l_\Omega$ and $W$ fits. The last column represents the $l_0$ inputted in the Monte-Carlo simulations.

Supplementary Note 1: Justification to consider the vortex as a particle in a 2D potential

The purpose of this section is to show that the vortex can be considered as a particle in a 2D potential landscape. We want to exclude, for example, that the vortex could stretch or bend, which would add complexity to our data analysis and interpretation.

Let's consider a vortex displacement is $l_\Omega$ parallel to the surface and a sample thickness is $d = 55$ nm. For simplicity, we assume this displacement results from a vortex tilt, as illustrated below. We calculate the energy difference $\Delta E$ necessary to tilt the vortex rather than keeping it straight. A more accurate vortex line would be obtained by imposing that the vortex leave the sample surfaces perpendicular to them. Notably, such a vortex line will be longer and thus require more energy.

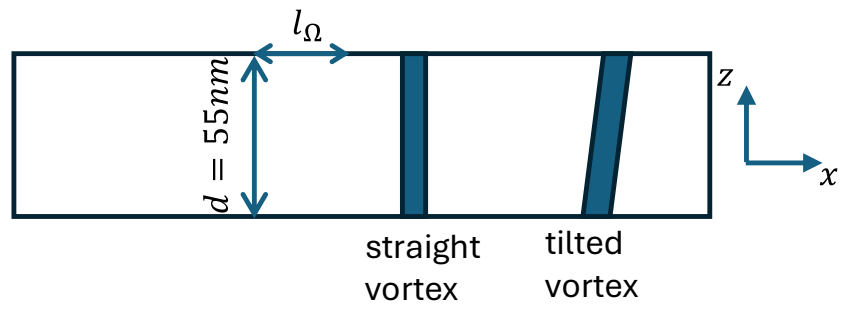

We consider the kinetic and magnetic energy of the vortex, neglecting the condensation energy. We use the following expression for the line energy per unit vortex length [1]

$$E_{line} = \frac{\Phi_0^2 l}{4\pi\mu_0\lambda^2}\ln\frac{\lambda}{\xi}$$

Where, $l$ is the length of the vortex, $\xi$ is the superconducting coherence length and $\lambda$ the London penetration depth. The additional energy cost to tilt the vortex is therefore:

$$\Delta E = \frac{\Phi_0^2}{4\pi\mu_0\lambda^2}\ln\frac{\lambda}{\xi}\left(d - \sqrt{d^2 + l_\Omega^2}\right)$$

Estimating $\xi$

The coherence length $\xi$ is estimated from the upper critical field at 4.2 K , $H_{c2}(4.2)$=2.25 T (see Supplementary Figure 7).

$$\xi(4.2\ k) = \sqrt{\frac{\phi_0}{2\pi H_{c2}(4.2)}} = 12\text{ nm}$$

Estimating $\lambda$

We estimate $\lambda$ by measuring the Pearl length from our SOT DC images $B_{DC}(x, y)$. We fit the images as we did in our previous work[2] and shown in Supplementary Figure 10. The measured Pearl length for our Nb film was 1.7 µm. From $\Lambda = 2\lambda^2/d$ we obtain that $\lambda = 220$ nm. The obtained value of $\lambda$ is comparable to other values quoted in the literature for 55 nm-thick films deposited by dc magnetron sputtering.[3] The resulting thermodynamic critical field is 90 mT (see equation below), which is slightly lower than expected for Nb $H_c^{th}(4.2) \approx 150$ mT.[4]

$$H_c^{th} = \frac{\Phi_0}{2\sqrt{2}\pi\lambda(T)\xi(T)} = 90\text{ mT}$$

Conclusion

Substituting $\lambda = 220$ nm, $\xi = 12$ nm, and $d = 55$ nm in the expression for $\Delta E$ we calculate the energy difference between the tilted vortex and the straight vortex for the cases of interest. For double well #3 and 4, $l_\Omega = 18$ nm in the strong drive limit, we obtain $\Delta E = 5 \times 10^{-20}$ which is more than an orders of magnitude larger then the energy scales we found, such as the energy barrier $E_b = 1 \times 10^{-21}$ J and the measured work $W < 2 \times 10^{-21}$ J. For double-well #1, the relevant energy scale are significantly larger with $E_b = 7 \times 10^{-21}$ J and $W = 3 \times 10^{-20}$ J. However, the displacement is also larger $l_\Omega = 50$ nm, which results in an energy cost larger by at least an order of magnitude $\Delta E = 4 \times 10^{-19}$ J. Therefore, we conclude that we can consider the vortex as a particle in a two-dimensional potential.

Supplementary Note 2: Estimation of the SQUID-on-tip force on the vortex

The magnetic field generated by the vortex and the magnetic field emitted by the SQUID-on-tip (SOT) loop lead to a vortex-tip interaction. Here, we experimentally estimate the force exerted by the SOT on the vortex.

We positioned the SOT near one of the two wells of the double-well potential (labeled well 1). We balance the wells by tuning the DC tilt current $I_{DC}$ until the vortex exhibits equal occupation between the two wells, as determined from the symmetric telegraphic noise signal. The $I_{DC}$ that balances the double-well is recorded ($I_{DC}^{+}$) for further comparison. We then reversed the sign of the voltage bias on the SOT, thereby changing the interaction from attractive/repulsive to repulsive/attractive. We repeat the same procedure to determine the new corresponding $I_{DC}$ value that balances the double-well. The difference in the resulting forces generated by the two distinct values of $I_{DC}$ required to balance the double-well gives us an estimate of the tip-vortex interaction. Formally, the force exerted by the SOT on the vortex was estimated using the following relations:

$$\Delta F_{DC} = \phi_0 \Delta J, \qquad \Delta J = \frac{\Delta I_{DC}}{\pi * \sqrt{\left(\frac{w_b}{2}\right)^2 - (x_{bc})^2}}, \qquad \Delta I_{DC} = \frac{I_{DC}^{+} - I_{DC}^{-}}{2}$$

Where $w_b$ is the width of the bridge, $x_{bc}$ is the distance between the vortex and the bridge center, and $I_{dc}^{+/-}$ are the DC currents required to balance the telegraphic noise when the tip–vortex interaction is attractive and repulsive, respectively. The expression for the sheet current density $J$ is taken from Ref. [5].

We repeated this procedure by locating the tip near the other (well 2) and repeated this measurement at three different tip–sample distances: $200$ nm, $300$ nm, and $400$ nm. The results, presented in Supplementary Figure 9, show the expected trend: the force required to balance the wells decreases with the tip–sample distance, consistent with the reduction of the magnetic coupling strength with distance. The interaction force at the scanning distance ($300$ nm) is comparable to the smallest driving force $F_{\Omega}$. However, it can be partially compensated by applying an external DC current $I_{DC}$.

We also observe minor discrepancies between the forces measured at the same height over well 1 and well 2. These differences likely arise from slight offsets in the lateral $(x, y)$ positioning of the tip relative to each well. This gives us an idea of the uncertainty in our estimate of the interaction force.

Supplementary Note 3: Phase Correction Procedure

The vortex is driven with a sinusoidal current. Due to low-pass filters and stray inductance and capacitance, a small phase is acquired between the room temperature voltage source and the current flowing in the sample. In general, this phase delay is driving frequency-dependent $\Omega$. Knowing the phase delay precisely is crucial as our estimation of the work is proportional to the out-of-phase component. To ensure accurate separation between the magnetic signal $B_z(t)$ components that are in-phase (ip) and out-of-phase (oop) with respect to the applied current, we search for the phase reference before vortex measurements. The SQUID-on-tip (SOT) is scanned over the superconducting bridge in a vortex-free region while applying a large oscillating current (1.2 mA) at the relevant frequencies $\Omega$. The large current ensures a strong and well-defined magnetic signal as expected by the Biot-Savart law. This magnetic signal is in-phase with the current flow and serves as our phase reference.

A lock-in amplifier (LIA) is used to simultaneously demodulate the in-phase ($B_\Omega^{ip}$) and out-of-phase ($B_\Omega^{oop}$) components of the first-harmonic magnetic field at each driving frequency $\Omega$. We then manually adjust the LIA's reference phase at each frequency so that the measured $B_\Omega^{oop}$ component is minimized and brought to the system's noise level. Supplementary Figure 1a shows $B_\Omega^{ip}(x)$ and $B_\Omega^{oop}(x)$ on the same scale, while Supplementary Figure 1b zooms in solely on $B_\Omega^{oop}(x)$. For optimal phase correction $\phi \equiv 0$, the $B_\Omega^{oop}(x)$ signal is at least 1000 times smaller than the $B_\Omega^{ip}(x)$. We also show how a small phase difference of $-0.16°$ is clearly visible. We record the precise phase settings for all relevant $\Omega$ and apply these phase corrections during subsequent measurements of vortices. This phase setting ensures that any oop signal detected reflects a lag in the vortex motion, rather than a lag acquired in the wiring.

Supplementary Note 4: Dissipation of a vortex in a harmonic potential

We consider an AC driving force $F_\Omega(t) = F_\Omega \cos \Omega t$. In this limit, viscous drag is small compared to the applied force, we know the vortex speed as a function of time $v(t) = \Omega l_\Omega \sin \Omega t$, where $l_\Omega$ is the vortex displacement amplitude. The low viscous drag approximation results in a higher speed estimate. Therefore, the work computed here is an upper bound. We assume the Bardeen-Stephen viscous drag[6] as the sole dissipation mechanism, $F_d = -\eta_0 v d$, where $\eta_0$ is the viscous drag per unit length and $d$ the film thickness. Thus, the work per cycle in this limiting case is given by $W = \pi d \eta_0 \Omega l_\Omega^2$.

According to Bardeen-Stephen $\eta_0 = H_c \Phi_0 / \rho_n$, where $H_c$ is the critical field and $\rho_n$ is the normal state resistance. For our Nb thin film, we measured the critical field $H_c(4.2\ \mathrm{K}) = 2.3$ T and normal state resistivity $\rho_n \sim 0.05\ \mu\Omega\mathrm{m}$ (Supplementary Figure 7). This results in $\eta_0 = 9 \times 10^{-8}\ \mathrm{Kgm^{-1}s^{-1}}$. In our Nb films, the vortex oscillation amplitude is $l_\Omega \approx 5$ nm. We consider the highest driving frequency $\Omega/2\pi = 150$ kHz. As a result, we estimate the vortex dissipation for a harmonic displacement to be $W \sim 10^{-25}$ J per cycle, which is significantly smaller than our smallest detected work $\sim 10^{-24}$J. We conclude that a single-well structure would not produce a detectable dissipation signal $l_\Omega^{ip}$.

Supplementary Note 5: Effect of Potential Tilt on Work and Displacement

To control the tilt between the two wells, we apply a static (DC) current $I_{DC}$. The optimal value of $I_{DC}$ is selected by maximizing both the in-phase and out-of-phase components of the first harmonic components while minimizing the second harmonic contribution, ensuring measurements are performed in the most symmetric (balanced) configuration. In this state, the critical force $F_c$ is minimized.

To verify the robustness of our method with respect to the choice of $I_{DC}$, we performed measurements under intentionally tilted conditions, where $I_{DC}$ was tuned to not fully balance the wells. Supplementary Figure 8 shows the work and displacement amplitude $l_\Omega = \sqrt{\left(l_\Omega^{ip}\right)^2 + \left(l_\Omega^{oop}\right)^2}$ as functions of the driving amplitude $F_\Omega$ for three different static tilt values—one tilt value we define as $F_{DC} \equiv 0$, which corresponds to the most symmetric potential. As expected, varying the tilt leads to a shift in the critical force $F_c$; stronger tilts increase $F_c$, consistent with the increased asymmetry of the potential.

Notably, we find that the relevant features of both $l_\Omega$ and the work remain unchanged across different small tilt values. In particular, below $F_c$, $l_\Omega$ grows linearly with $F_\Omega$, reflecting the harmonic trapping of the vortex. Above $F_c$, $l_\Omega$ saturates as deterministic hopping between wells occurs, and the system re-enters a linear response regime where the barrier is negligible. In this deterministic regime, both $l_\Omega$ and the work converge to the same values and are insensitive to the applied tilt. Thus, while $I_{DC}$ influences the value of $F_c$, it does not significantly modify the underlying vortex response for $F_\Omega < F_c$ or $F_\Omega > F_c$, provided that the tilt is sufficiently small not to lock the vortex into one of the wells.

For a large tilt, the vortex we expect the signature of a vortex trapped in a harmonic potential. To test this assumption, we measured the vortex displacement amplitude $l_\Omega$ and work $W$ as functions of frequency in double-well potentials under two tilt conditions: balanced wells (no tilt, $F_{DC} \equiv 0$) and fully tilted ($F_{DC} = 243$ fN). The vortex is driven with an oscillating driving force amplitude $F_\Omega = 188$ fN for all driving frequencies $\Omega$. The results are shown in Supplementary Figure 4. For the fully tilted case, the vortex displacement does not depend on frequency, consistent with being confined in a harmonic potential well. In contrast, the dissipative (no tilt) state shows frequency-dependent work (Supplementary Figure 4a), while the tilted state, with the vortex confined to one well, shows vanishing dissipation across frequencies. This highlights the difference between a vortex that permanently oscillates within one harmonic potential and one that hops over the barrier between two wells.

Supplementary Note 6: Maximum switching frequency

In this note, we estimate the maximum switching frequency of the vortices trapped in double-wells #1,2, and 3.

We consider that the maximum switching frequency of a memory bit based on our vortex is limited by the shuttling time of the vortex between two states. To estimate this time-of-flight, we assume that the vortex energy relaxes entirely through the Bardeen-Stephen viscous drag. Thus, the dissipation over one cycle (two switches) is $W = \oint \vec{F} \cdot d\vec{x} = -4\eta_0 \overline{v} d l_\Omega$, where $\eta_0 = H_c \Phi_0 / \rho_n$ is the viscous drag per unit length, $\overline{v}$ is the average speed, $d$ is the film thickness, and $l_\Omega$ is the oscillation amplitude. For our Nb thin film, we measured the critical field $H_c(4.2\ \mathrm{K}) = 2.3$ T and normal state resistivity $\rho_n \sim 0.05\ \mu\Omega\mathrm{m}$ (Supplementary Figure 7). This results in $\eta_0 = 9 \times 10^{-8}\ \mathrm{Kgm^{-1}s^{-1}}$.

Considering double-well #3, we know the value for the maximum work and displacement at the maximum driving amplitude $W = 3 \times 10^{-21}$ J, $l_\Omega = 10$ nm, we obtain $\overline{v} = 5$ m/s. Such speed is relatively slow compared to what is obtained in the flux flow regime ~10 km/s[7] where the Bardeen-Stephen viscous drag was found to be a good approximation. At such an average speed, the time to switch the vortex state is $\tau = 4$ ns, which implies switching frequencies of $250$ MHz.

Higher operating frequencies can be achieved by utilizing better materials and pinning landscape engineering. The maximum operating frequencies of double-wells #1 and 2 are 300 and 833 MHz, respectively.

A similar argument is made in the context of finite-Landauer's bound[8], where the work penalty has an additional term that goes like $kT\,\tau_0/2\tau$, where $\tau_0 = \frac{\eta_0 d l_0^2}{k_b T} = 8 \times 10^{-9}$ s. This also supports that the vortex could be operated at such frequencies. Reducing further $\tau_0$ could be done by using other materials such as YBCO, which has a smaller coherence length and thus smaller $l_0$, larger $H_c$, and higher normal state resistance, which translates into lower viscosity $\eta_0$. In such material, the pinning potential could be engineered using a helium focused ion beam[9].